\documentclass[a4paper]{article}

\usepackage{geometry}
\usepackage{subfigure}
\usepackage{epsfig}
\usepackage{url}
\usepackage{amsfonts}
\usepackage{tipa}
\usepackage{times}
\usepackage{mathrsfs}
\usepackage{amssymb}
\usepackage{amsthm}
\usepackage{amsmath}

\newtheorem{definition}{Definition}[section]
\newtheorem{lemma}{Lemma}[section]
\newtheorem{theorem}{Theorem}[section]

\newtheorem{fact}{Fact}[section]

\newcommand{\Var}{\textrm{Var}}
\newcommand{\vol}{\textrm{vol}}

\newcommand{\poly}{\textrm{poly}}

\begin{document}

\title{Structure Entropy and Resistor Graphs
\footnote{The authors are partially supported by the Grand Project
``Network Algorithms and Digital Information'' of the Institute of
Software, Chinese Academy of Sciences, by an NSFC grant No.
61161130530, and by a China Basic Research Program (973) Grant No. 2014CB340302.}}

\author{Angsheng Li\\
State Key Laboratory of Computer Science\\
 Institute of Software,
 Chinese Academy of Sciences\\
 School of Computer Science, University of Chinese Academy of Sciences\\ Beijing, 100190, P. R. China\\  
Yicheng Pan\\
State Key Laboratory of Computer Science\\
 Institute of Software,
 Chinese Academy of Sciences\\
 Beijing, 100190, P. R. China}

\maketitle

\begin{abstract}
The authors \cite{LP2016a} defined the notion of structure entropy of a graph $G$ to measure the information embedded in $G$ that determines and decodes the essential structure of $G$. Here, we propose the notion of {\it resistance of a graph} as an accompanying notion of the structure entropy to measure the force of the graph to resist cascading failure of strategic virus attacks.
We show that for any connected network $G$, the resistance of $G$ is $\mathcal{R}(G)=\mathcal{H}^1(G)-\mathcal{H}^2(G)$, where $\mathcal{H}^1(G)$ and $\mathcal{H}^2(G)$ are the one- and two-dimensional structure entropy of $G$, respectively.
According to this, we define the notion of {\it security index of a graph} to be the normalized resistance, that is, $\theta (G)=\frac{\mathcal{R}(G)}{\mathcal{H}^1(H)}$.
We say that a connected graph is an $(n,\theta)$-{\it resistor graph}, if $G$ has $n$ vertices and has security index $\theta (G)\geq\theta$. We show that trees and grid graphs are $(n,\theta)$-resistor graphs for large constant $\theta$, that the graphs with bounded degree $d$ and $n$ vertices, are $(n,\frac{2}{d}-o(1))$-resistor graphs, and that for a graph $G$ generated by the security model \cite{LLPZ2015, LP2016}, with high probability, $G$ is an $(n,\theta)$-resistor graph, for a constant $\theta$ arbitrarily close to $1$, provided that $n$ is sufficiently large. To the opposite side, we show that expander graphs are not good resistor graphs, in the sense that, there is a global constant $\theta_0<1$ such that expander graphs cannot be $(n,\theta)$-resistor graph for any $\theta\geq\theta_0$. In particular, for the complete graph $G$, the resistance of $G$ is a constant $O(1)$, and hence the security index of $G$ is $\theta (G)=o(1)$. This shows that, for arbitrarily small constant $\epsilon>0$, there is an $N$ such that for any $n\geq N$, the complete graph $G$ of $n$ vertices cannot be an $(n,\epsilon)$-resistor graph. Finally, we show that for any simple and connected graph $G$, if $G$ is an $(n, 1-o(1))$-resistor graph, then there is a large $k$ such that the $k$-th largest eigenvalue of the Laplacian of $G$ is $o(1)$, giving rise to an algebraic characterization for the graphs that are secure against intentional virus attack.

\end{abstract}

\section{\bf Introduction}\label{sec:int}

Shannon's \cite{S1948} metric measures the uncertainty of a probabilistic distribution as

\begin{equation}
H(p_1,\cdots,p_n)=-\sum\limits_{i=1}^np_i\log_2p_i.
\end{equation}
This metric and the associated concept of noise, have provided rich sources for both information science and technology.
However, as pointed out by Brooks in \cite{B2003}, it had been a longstanding challenge to define the information that is embedded in a physical system, which determines and decodes the essential structure of the (observed and noisy) physical system. Such a metric, if well defined, may provide an approach to understand the folded three-dimensional structures of proteins.
Shannon \cite{S1953} himself realized that his metric of information fails to support the analysis of communication networks to answer the question such as characterization of the optimal communication networks. The answer for this question depends on a well-defined definition of the structure entropy, that is, the information embedded in a communication network.

The challenge of the quantification of structural information becomes more and more important in the current information age, in which noisy big data with or without structures are assumed to support the world and our societies. Structural information may provide the principles for structuring the unstructured data and for discovering the knowledge from noisy data by removing the noises.
To this end, the authors \cite{LP2016a} introduced the notion of coding tree of a graph, and defined the structure entropy of a graph to be the minimum amount of information required to determine the codeword of the vertex in a coding tree for the vertex that is accessible from random walk with stationary distribution in the graph. The structure entropy of a graph is hence the information embedded in the graph. The structural information of a graph defined in this way allows us to decode the essential structure of the graph simultaneously at the same time when we measure the structure entropy of the graph.

In the present paper, we analyze communication networks based on the structural information theory \cite{LP2016a}. Specifically, we investigate the security of networks against cascading failure of virus attacks. We
introduce the notion of {\it resistance of a graph} and {\it resistor graphs} as accompanying notions of structure entropy of graphs to analyze the security of networks.

Network security has become a grand challenge in modern information science and computer science.
An interesting discovery in network theory in the last few years is
that network topology is universal in nature, society, and
industry~\cite{Bar2009}. In fact, the current highly connected world
is assumed to be supported by numerous networking systems. Real
networks are not only too important to fail, but also too
complicated to understand.

Erd\"os-R\'enyi proposed the first model \cite{ER1959, ER1960} (The
ER model in short) to capture complex systems based on the
assumption that real systems are evolved randomly. The ER model explores that if a graph is generated randomly, then the diameter of the graph is exponentially smaller than the size of the graph, referred to as the {\it small world phenomenon}.
It has been shown that most real world networks do satisfy the small world phenomenon, giving rise to the first universal property of networks.
However, can we really assume that real networks are purely random?
Barab\'asi and Albert \cite{Bar1999} proposed a graph generator by introducing preferential attachment as an explicit mechanism, the model is thus called the preferential
attachment (PA) model. Consequently, networks generated by the PA model naturally
follow a power law. It has been shown that most real networks follow
a {\it power law}. Consequently, power law has become the second universal property of
networks \cite{Bar2009}.

 As a matter of fact, real world networks are highly connected and naturally evolving, in which information
spread easily and quickly. This is of course one of the main advantages of networks in both theory and applications.
However, at the same time, this could be one of the main disadvantages of networks. Because, virus also quickly spreads all over the networks.
It is due to this reason that network security has become a grand new challenge in the 21st century.

Networks may fail under attack due to different mechanisms \cite{AM1991, M2000, KKT2003, KKT2005, SFSRVR2009,
AJB2000}. The first type is the physical
attack of removal of  some vertices or edges. It has been shown that in
scale-free networks of the preferential attachment (PA)
model~\cite{Bar1999}, the overall network connectivity measured by
the sizes of the giant connected components and the diameters does
not change significantly under random removal of a small fraction of
vertices, but vulnerable to removal of a small fraction of the high-degree vertices~\cite{AJB2000, CRB2001, M2004}. The second type is the
cascading failure of attacks, which naturally appeared in rumor
spreading, disease spreading, voting, and advertising~\cite{W2002,AM1991,M2000}. One of the main features of networks
in the current highly connected world is that failure of a few vertices
of a network may generate a cascading failure of the whole network. It has been shown
that in scale-free networks of the preferential attachment model
even a weakly virulent virus can spread ~\cite{PV2001}. This explains
a fundamental characteristic of security of
networks~\cite{SFSRVR2009}.

For the physical attacks or random errors of removal of vertices, it was shown that the optimal networks resisting both physical attacks and random errors have at most
three values of degrees for all the vertices of the networks \cite{AVSS2004}, that networks having the optimal robustness resisting both high-degree vertices attacks and random errors, has a bimodal degree distribution \cite{TPCHS2005}.
To enhance the robustness of networks against biological virus spreading, it was proposed in \cite{CHA2003} the acquaintance immunization strategy, which calls for the immunization of random acquaintances of randomly chosen vertices, and more recently, a security enhancing algorithm
was proposed in \cite{SMA+2008} by randomly swapping two edges for a number of pairs of edges.

Li et al \cite{LLPZ2015} proposed the security model of networks by using the idea of the Art of War \cite{SUN2007}. It has been shown that with appropriate choices of parameters, the networks generated by the security model are secure against attacks of small scales \cite{LP2016}. Li and Pan \cite{LP2016a} proposed the notion of structure entropy of networks to quantitatively measure the dynamical complexity of interactions and communications of the network, for each natural number $K$.
However, it is an important open question to define a measure of security of a graph against cascading failure of intentional virus attacks. The authors of this paper and his coauthors \cite{LHLP2016} proposed the notions of resistance and security index of a graph by using the one- and two-dimensional structure entropy, and verified that both the resistance and security index measure the force of the graph to resist cascading failure of virus attacks. In \cite{LZP2016}, it was shown that resistance maximization is in fact the principle for defending networks against the super virus that infect all the neighbor vertices immediately. 
In the present paper, we establish the basic theory of resistance and security index of graphs.

We organise the paper as follows. In Section \ref{sec:structuralentropy}, we introduce and analyze the notions of coding tree and structure entropy proposed by the authors in \cite{LP2016a}, we also introduce a variation of the structure entropy to study the relationship between the Shannon entropy and the structure entropy.
In Section \ref{sec:resistance}, we define the notions of resistance and security index of networks, and establish both the local and global resistance laws of networks.
In Section \ref{sec:basic}, we introduce some basic results of the resistance and security indices of networks. In Section \ref{sec:bounded degree graph}, we establish a lower bound of the resistance of bounded degree graphs. In Section \ref{sec:complete graph}, we show that the resistance of a complete graph is actually a universal constant $O(1)$.
In Section \ref{sec:security}, we establish the theory of the resistance and security indices of the networks generated by the security model. In Section \ref{sec:strong resistor graph}, we establish both the combinatorial and algebraic characterization theorems for the graphs with the optimal two-dimensional structure entropy, i.e., $\mathcal{H}^2(G)=O(\log_2\log_2n)$. In Section \ref{sec:resistor graph}, we establish both the combinatorial and the algebraic characterization theorems for the resistor graphs.
In Section \ref{sec:con}, we summarise the results of the paper.

\section{Structure Entropy of Graphs}\label{sec:structuralentropy}

The authors of this paper \cite{LP2016a} proposed the notion of structure entropy of a graph to measure the information embedded in a physical system that decodes the essential structure of the system.
In this section, we introduce the notion of structure entropy.

Before introducing the structure entropy, we recall the Huffman codes \cite{H1952}.

\subsection{Huffman codes}

Suppose that $\Sigma=\{1, 2,\cdots, n\}$ such that the probability $i$ occurs is $p_i$, for each $i$. Let $\sum\limits_{i=1}^np_i=1$. We will encode the elements of $\Sigma$ by $0,1$-strings such that there is no codeword of an element is an initial segment of the codeword of another element of $\Sigma$.

Suppose that $T$ is a binary tree whose leaves are the codewords of the elements $1,2,\cdots, n$. Suppose that element $i$ has probability $p_i$ with codeword $\alpha_i$ in $T$, then the average length of the codewords is

\begin{equation}\label{eqn:Huffman-0}
L^T(p_1,p_2,\cdots,p_n)=\sum\limits_{i=1}^np_i\cdot |\alpha_i|,
\end{equation}
where $|\alpha_i|$ is the length of $0,1$-string $\alpha_i$.

The Huffman codes are to find the binary tree $T$ such that $L^T(p_1,p_2,\cdots,p_n)$ in Equation (\ref{eqn:Huffman-0}) is minimized.

We define

\begin{equation}
L(p_1, p_2, \cdots, p_n)=\min_TL^T(p_1, p_2, \cdots, p_n),
\end{equation}
where $T$ ranges over all the binary trees of $n$ leaves.

By definition, $L(p_1, p_2, \cdots, p_n)$ is the minimum average length of the binary representation of the alphabet $\Sigma$. Huffman codes achieve the minimum solution $L(p_1, p_2, \cdots, p_n)$.

It is known that

\begin{equation}
L(p_1, p_2, \cdots, p_n)\geq H(p_1, p_2, \cdots, p_n),
\end{equation}
with equality holds when $p_i=2^{-k}$ for some $k$, and for all $i$, where $H(p_1, p_2, \cdots, p_n)$ is the Shannon entropy of ${\bf p}=(p_1, p_2, \cdots, p_n)$.

This means that the minimum average length of the binary representation of an element picked from a probability distribution is lower bounded by the Shannon entropy of the distribution, and the Shannon entropy is the tight lower bound of the minimum average length of the binary representations.

Before developing our theory, we recall a basic interpretation of the $\log_2$ function:

Let $p$ be a number with $0<p<1$. Suppose that $k$ is a natural number such that

\begin{equation}
\frac{1}{2^{k+1}}\leq p<\frac{1}{2^k},
\end{equation}
which implies that

\begin{equation}\label{eqn:log}
k+1\geq -\log_2p>k.
\end{equation}

Equation (\ref{eqn:log}) indicates that

\begin{enumerate}
\item [(i)] $-\log_2p$ is the information (or uncertainty) embedded in an item that occurs with probability $p$.
\item [(ii)] $\lceil -\log_2p\rceil$ many bits are sufficient to express the item that occurs with probability $p$.
\item [(iii)] The minimum length of the binary codeword of the item that occurs with probability $p$ is exactly $\lceil -\log_2p\rceil$, which is greater than or equal to the information embedded in the item occurring with probability $p$, that is, $-\log_2p$.
\end{enumerate}

To define the structure entropy of a graph, we first need to encode a graph. Similarly to the Huffman codes, we encode a graph by a tree. However, it is a priority tree below, instead of a binary tree in the Huffman codes.

\subsection{Priority tree}

\begin{definition} (Priority tree) A priority tree is a rooted tree $T$ with the following properties:

\begin{enumerate}
\item [(i)] The root node is the empty string, written $\lambda$.

A node in $T$ is expressed by the string of the labels of the edges from the root to the node.
We also use $T$ to denote the set of the strings of the nodes in $T$.

\item [(ii)] For every node $\alpha$ in $T$, there is a natural number $k$ such that there are $k$ edges linking $\alpha$ to its $k$ children. The edges are labelled by

$$0<1<\cdots <k-1.$$

({\it Remark: (i)
Unlike Huffman codes, we use an alphabet of the form $\Sigma=\{0,1,\cdots, k\}$ for each tree node $\alpha$. In the Huffman codes, we always use the alphabet $\Sigma=\{0,1\}$.

(ii) Different nodes in $T$ may have different numbers of children, i.e., different $k$'s.)}

\item [(iii)] Every tree node $\alpha$ is a string of numbers from $0$ to some natural number.

\end{enumerate}

\end{definition}

For two tree nodes $\alpha, \beta$, if $\alpha$ is an initial segment of $\beta$ as strings, then we write $\alpha\subseteq\beta$. If $\alpha\subseteq\beta$ and $\alpha\not=\beta$, we write $\alpha\subset\beta$.

\subsection{Coding tree of a graph}

\begin{definition} (Coding tree of a graph) Let $G=(V,E)$ be a graph. A coding tree of $G$ is a priority tree $T$ such that every tree node $\alpha\in T$, there is a subset $T_{\alpha}$ of the vertices $V$, and such that the following properties hold:

\begin{enumerate}
\item [(i)] The root node $\lambda$ is associated with the whole set $V$ of vertices of $G$, that is, $T_{\lambda}=V$.
\item [(ii)] For every node $\alpha\in T$, if $\beta_1,\beta_2,\cdots,\beta_k$ are all the children of $\alpha$, then $\{T_{\beta_1},\cdots, T_{\beta_k}\}$ is a partition of $T_{\alpha}$.
\item [(iii)] For every leaf node $\gamma\in T$, $T_{\gamma}$ is a singleton.

\end{enumerate}

\end{definition}

\begin{definition} (Codeword) Let $G=(V,E)$ be a graph, and $T$ be a coding tree of $G$.

\begin{enumerate}
\item [(i)] For every node $\alpha\in T$, we call $\alpha$ the codeword of set $T_{\alpha}$, and $T_{\alpha}$ the marker of $\alpha$.
\item [(ii)] For a leaf node $\gamma\in T$, if $T_{\gamma}=\{v\}$, then we say that $\gamma$ is the codeword of $v$, and $v$ is the marker of $\gamma$.
\end{enumerate}

\end{definition}

A coding tree $T$ of a graph $G$ satisfies the following

\begin{definition}\label{def:coding tree property}
(Coding tree properties) Given a graph $G$ and a coding tree $T$ of $G$, we assume that the following properties hold:

\begin{enumerate}
\item [(i)] For every node $\alpha\in T$, the marker $T_{\alpha}$ of $\alpha$ is explicitly determined. This means that if we know $\alpha$, then we have already known the marker $T_{\alpha}$. This means that there is no uncertainty in $T_{\alpha}$ once we know the codeword $\alpha\in T$.

    \item [(ii)] For every node $\alpha\in T$, if we know the codeword $\alpha$, then we simultaneously know $\beta$ for all the codewords $\beta$'s in the branch between the root node $\lambda$ and $\alpha$ in $T$, i.e., the $\beta$ with $\beta\subseteq\alpha$.

\end{enumerate}
\end{definition}
The advantage of the coding tree is the coding tree properties in Definition \ref{def:coding tree property}. The key to our definition of structure entropy is to use the coding tree properties above to reduce the uncertainty of a graph by a coding tree of the graph.

\begin{lemma} \label{lem:codeword} Let $G=(V,E)$ be a graph and $T$ be a coding tree of $G$. Then:

\begin{enumerate}
\item [(1)] For every leaf node $\gamma\in T$, there is a unique vertex $v$ such that $\gamma$ is the codeword of $v$.
\item [(2)] For every vertex $v\in V$, there is a unique leaf node $\gamma\in T$ such that $v$ is the marker of $\gamma$.

\end{enumerate}

\end{lemma}
\begin{proof} By the definition of coding tree.
\end{proof}

By Lemma \ref{lem:codeword}, the set of all the leaves in $T$ is the set of codewords of the vertices $V$. This property is the same as the Huffman codes, that is, all the leaf nodes of the tree are the codewords desired.

\subsection{Structure entropy of a graph given by a coding tree}

The authors \cite{LP2016a} introduced the notion of structure entropy of a graph.

\begin{definition} \label{def:structure entropy-T-O} (Structure entropy of a graph by a coding tree, Li and Pan \cite{LP2016a}) Let $G=(V,E)$ be a graph, and $T$ be a coding tree of $G$.
We define the structure entropy of $G$ by coding tree $T$ as follows:

\begin{equation}\label{eqn:H-T-O}
\mathcal{H}^{T}(G)=-\sum\limits_{\alpha\not=\lambda, \alpha\in\ T}\frac{g_{\alpha}}{{\rm vol}(G)}\cdot \log_2\frac{{\rm vol}(\alpha)}{{\rm vol}(\alpha^{-})},
\end{equation}
where $g_{\alpha}=|E(\bar{T_{\alpha}}, T_{\alpha})|$, that is, the number of edges from the complement of $T_{\alpha}$, i.e., $\bar{T_{\alpha}}$ to $T_{\alpha}$, ${\rm vol}(G)$ is the volume of $G$, that is, the total degree of vertices in $G$, ${\rm vol}(\beta)$ is the volume of the vertices set $T_{\beta}$, and $\alpha^{-}$ is the parent node of $\alpha$ in $T$.

\end{definition}

To understand Equation (\ref{eqn:H-T-O}), we conclude the following items for the metric $\mathcal{H}^T(G)$:

\begin{enumerate}
\item [(1)] For every node $\alpha\in T$, $T_{\alpha}$ is the set of vertices associated with $\alpha$.
Suppose that, once we known $\alpha$, we have already known the set $T_{\alpha}$.

\item [(2)] For each node $\alpha\in T$ with $\alpha\not=\lambda$, since $\alpha^{-}$ is the parent node of $\alpha$ in $T$, the probability that the vertex $v\in V$ from random walk with stationary distribution in $G$ is in $T_{\alpha}$ under the condition that $v\in T_{\alpha^{-}}$ is $\frac{{\rm vol}(\alpha)}{{\rm vol}(\alpha^{-})}$. Therefore the entropy (or uncertainty) of $v\in T_{\alpha}$ under the condition that $v\in T_{\alpha^{-}}$ is $-\log_2\frac{{\rm vol}(\alpha)}{{\rm vol}(\alpha^{-})}$.

\item [(3)] For every node $\alpha\in T$, $g_{\alpha}$ is the number of edges that random walk with stationary distribution arrives at $T_{\alpha}$ from vertices $\bar{T_{\alpha}}$, the vertices outside $T_{\alpha}$. Therefore, the probability that a random walk with stationary distribution is from outside $T_{\alpha}$ to vertex in $T_{\alpha}$ is $\frac{g_{\alpha}}{{\rm vol} (G)}$.

\end{enumerate}

 Consider the stochastic process of random walks with stationary distribution in $G$. It is the stochastic process as follows:
\begin{equation}\label{eqn:sp}
X_0, X_1, X_2,\cdots,
\end{equation}
with the following properties:
\begin{enumerate}
\item [(i)] Let $x_0$ be the vertex chosen in $V$ with probability proportional to vertex degree, and $X_0$ be the codeword of vertex $x_0$ in $T$.
Suppose that $X_i$ and $x_i$ are defined.
\item [(ii)] Let $x_{i+1}$ be the neighbor of $x_i$ chosen uniformly and randomly among all the neighbors of $x_i$ in $G$. Then $X_{i+1}$ is defined as the codeword of $x_{i+1}$ in $T$.

\end{enumerate}

In the stochastic process in Equation (\ref{eqn:sp}), we are interested in the quantification of the entropy of $X_{i+1}$ under the condition that we have already known $X_i$, denoted by

\begin{equation}\label{eqn:random walk}
\widetilde{H}(X_{i+1}|X_i).
\end{equation}

To compute $\widetilde{H}(X_{i+1}|X_i)$, suppose that $x_i$ and $x_{i+1}$ are as above, and that $X_i=\alpha\in T$ is the codeword of $x_i$ in $T$. Notice that the codeword $\alpha$ is a leaf node in $T$. By Definition \ref{def:coding tree property}, we know $T_{\delta}$ for all the nodes $\delta\subseteq\alpha$, i.e., the initial segments of $\alpha$ as strings.

Let $\gamma$ be the longest node $\delta\in T$ with $\delta\subseteq\alpha$ such that $x_{i+1}\in T_{\delta}$ holds. Then we know that $\gamma$ is an initial segment of the codeword of $x_{i+1}$ in $T$. To determine the codeword of $x_{i+1}$ in $T$, we only need to find the branch from $\gamma$ to a leaf node $\beta\in T$ such that $x_{i+1}\in T_{\beta}$. According to the analysis above, the information of $X_{i+1}$ under the condition of $X_i$ is:

\begin{equation*}
\widetilde{H}(X_{i+1}=\beta|X_i=\alpha)=-\sum\limits_{\delta\in T,\ \gamma\subset\delta\subseteq\beta}\frac{g_{\delta}}{{\rm vol}(G))}\log_2\frac{{\rm vol}(\delta)}{{\rm vol}(\delta^{-})},
\end{equation*}
where $\gamma=\alpha\cap\beta$ is the node in $T$ at which $\alpha$ and $\beta$ branch in $T$, $g_{\delta}$ is the number of edges in the cut $(T_{\alpha},\bar{T_{\delta}})$.

We notice that, only if  both $x_{i+1}\in T_{\delta}$ and $x_i\not\in T_{\delta}$ occur, we need to determine the codeword of $T_{\delta}$ in $T_{\delta^{-}}$, for which the amount of information required is $-\log_2\frac{{\rm vol}(\delta)}{{\rm vol}(\delta^{-})}$. So, intuitively, $\widetilde{H}(X_{i+1}=\beta|X_i=\alpha)$ is the amount of information, in terms of the codeword of $T_{\delta}$ in $T_{\delta^{-}}$, required to determine the codeword of $x_{i+1}$ under the condition that the codeword of $x_i$ is known. Note that we use the codewords of nodes in the coding tree to measure amount of information. That is why we use the notation $\widetilde{H}(\cdot)$ to distinguish from the classic conditional entropy notation $H(\cdot)$.

Our definition of $\mathcal{H}^T(G)$ in Definition \ref{def:structure entropy-T-O} is
\begin{equation*}
\mathcal{H}^T(G)=\sum\limits_{\substack{e=(x_i,\ x_{i+1})\\ x_i, x_{i+1}\in V}}
\widetilde{H}(X_{i+1}=\beta|X_i=\alpha),
\end{equation*}
where $X_i$ is the codeword of $x_i$, and $X_{i+1}$ is the codeword of $x_{i+1}$.

This measures the information required to determine the codeword of the vertex in $V$ that is accessible from random walk with stationary distribution in $G$, under the condition that the codeword of the starting vertex of the random walk is known.

\subsection{Structure entropy}

\begin{definition}\label{def:structure-entropy-O} (Structure entropy of a graph, Li and Pan \cite{LP2016a}) Let $G=(V,E)$ be a graph.

\begin{enumerate}
\item [(1)] The structure entropy of $G$ is defined as

\begin{equation}
\mathcal{H}(G)=\min_{T}\{\mathcal{H}^T(G)\},
\end{equation}
where $T$ ranges over all the coding trees of $G$.

[{\it Remark: (i) We notice that the Huffman codes require to find a binary tree $T$ such that the $L^T(p_1,\cdots,p_n)$ in Equation (\ref{eqn:Huffman-0}) is minimized. In this case, Huffman codes have already been the optimum solution.

(ii) Our structure entropy of a graph requires to find a coding tree $T$ such that the $\mathcal{H}^T(G)$ in Equation (\ref{eqn:H-T-O}) is minimized.
However, there is no algorithm achieving the optimal structure entropy so far. Although there are nearly liner time greedy algorithms for approximating the optimum coding tree, with remarkable applications. \cite{LP2016a}
}]

\item [(2)] For natural number $K$, the $K$-dimensional structure entropy of $G$ is defined as

\begin{equation}
\mathcal{H}^K(G)=\min_{T}\{\mathcal{H}^T(G)\},
\end{equation}
where $T$ ranges over all the coding trees of $G$ of heights at most $K$.

[{\it Remark}: This allows us to study the structure entropy in different dimensions.]

\end{enumerate}

\end{definition}

The metric $\mathcal{H}(G)$ has the following intuitions:

\begin{itemize}
\item The structure entropy $\mathcal{H}(G)$ of $G$ is the least amount of information required to determine the codeword of the vertex in a coding tree that is accessible from random walk with stationary distribution in $G$.
\item The structure entropy $\mathcal{H}(G)$ is the information that determines and decodes the coding tree $T$ of $G$ that minimizes the uncertainty in positioning the vertex that is accessible from random walk in graph $G$.

Therefore, $\mathcal{H}(G)$ is not only the measure of information, but decodes the structure of $G$ that minimises the uncertainty in the communications in the graph, which can be regarded as the ``essential" structure of the graph.

\end{itemize}

The $K$-dimensional structure entropy $\mathcal{H}^K(G)$ of $G$ has the similar intuitions as above.

We remark that the structure entropy has a rich theory with remarkable applications, more details are referred to \cite{LP2016a}.
Here we develop the theory of security by using the structure entropy with the cut module function as defined in \cite{LP2016a}. It is interesting to notice that a theory of network security can be established by using only the one- and two-dimensional structure entropy developed in \cite{LP2016a}.

\subsection{A variation of structure entropy}

%\subsection{Module function of a graph}

The structure entropy of a graph in Definitions \ref{def:structure entropy-T-O} and \ref{def:structure-entropy-O} is often misunderstood as the average length of the codeword of the vertex that is accessible from random walk with stationary distribution in the graph. We argue that this is not the case.

To better understand the question, we introduce a variation of the structure entropy. It depends on a module function of a graph.

\begin{definition}\label{def:module function} (Module function) Let $G=(V,E)$ be a connected graph. Let ${\rm vol}(G)$ be the volume of $G$. A module function of $G$ is a function $g$ of the form:

\begin{equation}
g:\ 2^V\ \rightarrow \{0, 1, \cdots, {\rm vol}(G)\}.
\end{equation}
\end{definition}

We define the structure entropy of a graph with a module function as follows.

\begin{definition} \label{def:structure entropy-T} (Structure entropy of a graph with a module function by a coding tree) Let $G=(V,E)$ be a graph, $g$ be a module function of $G$, and $T$ be a coding tree of $G$.
We define the structure entropy of $G$ with module function $g$ by coding tree $T$ as follows:

\begin{equation}\label{eqn:H-T}
\mathcal{H}^{T}_g(G)=-\sum\limits_{\alpha\not=\lambda, \alpha\in\ T}\frac{g(T_\alpha)}{{\rm vol}(G)}\cdot \log_2\frac{{\rm vol}(\alpha)}{{\rm vol}(\alpha^{-})},
\end{equation}
where ${\rm vol}(G)$ is the volume of $G$, ${\rm vol}(\beta)$ is the volume of the vertices set $T_{\beta}$, and $\alpha^{-}$ is the parent node of $\alpha$ in $T$.

\end{definition}

%\subsection{Structure entropy with a module function}

\begin{definition}\label{def:structural-entropy} (Structure entropy of a graph with a module function) Let $G=(V,E)$ be a graph, and $g$ be a module function of $G$.

\begin{enumerate}
\item [(1)] The structure entropy of $G$ with module function $g$ is defined as

\begin{equation}
\mathcal{H}_g(G)=\min_{T}\{\mathcal{H}^T_g(G)\},
\end{equation}
where $T$ ranges over all the coding trees of $G$.

\item [(2)] For natural number $K$, the $K$-dimensional structure entropy of $G$ with module function $g$ is defined as

\begin{equation}
\mathcal{H}^K_g(G)=\min_{T}\{\mathcal{H}^T_g(G)\},
\end{equation}
where $T$ ranges over all the coding trees of $G$ of heights at most $K$.

[{\it Remark}: This allows us to study the structure entropy in different dimensions.]

\end{enumerate}

\end{definition}

The formula $\mathcal{H}^{T}_g(G)=-\sum\limits_{\alpha\not=\lambda, \alpha\in\ T}\frac{g(T_\alpha)}{{\rm vol}(G)}\cdot \log_2\frac{{\rm vol}(\alpha)}{{\rm vol}(\alpha^{-})}$ is a generalization of $\mathcal{H}^T(G)$ in Definition \ref{def:structure entropy-T-O} with the function $g$ here being an arbitrarily given module function, while the function $g$ in Definition \ref{def:structure entropy-T-O} is the cut module function, that is, the number of edges in the cut.

In Definitions \ref{def:structure entropy-T} and \ref{def:structural-entropy}, the structure entropy of graph $G$ depends on a choice of a module function $g$. It is possible that there are many interesting choices for the module function $g$. We list a few of these as example:

\begin{enumerate}
\item [(i)] For a subset $X$ of vertices $V$, $g(X)$ is the volume of $X$. In this case, $g$ is called the {\it volume module function}.
\item [(ii)] For each subset $X$ of $V$, $g(X)$ is the weights in the cut $(X,\bar{X})$ in $G$. In this case, we say that $g$ is the {\it cut module function}.
\item [(iii)] For a directed graph $G$ and for each subset $X$ of $V$, $g(X)$ is the weights of the flow from $\bar{X}$ to $X$. In this case, we call $g$ the {\it flow module function}.

For directed graphs, the flow module function would be essential to the structure entropy of the graphs.
\end{enumerate}

In particular, there are module functions with additivity, with which the structure entropy collapses to the case of Shannon entropy.

\begin{definition}\label{def:additive function} (Additive module function) Let $G=(V,E)$ be a connected, simple graph with $n$ vertices and $m$ edges, and $g$ be a module function of $G$.
 We say that $g$ is an additive module function if for any disjoint sets $X$ and $Y$ of $V$,

\begin{equation}
g(X\cup Y)=g(X)+g(Y).
\end{equation}

\end{definition}

\begin{theorem} \label{thm:additive function theorem} (Structural entropy of a graph with an additive function) Let $G=(V,E)$ be a connected, simple graph with $n$ vertices, and $m$ edges, and let $g$ be an additive module function of $G$. For any coding tree $T$ of $g$, if $g$ satisfies the boundary condition
	$$g(T_\alpha)=
	\begin{cases}
	d_\alpha & \text{\rm $\alpha$ is a leaf}\\
	2m & \text{$\alpha=\lambda$}
	\end{cases},$$	
then
\begin{equation}
\mathcal{H}^T_g(G)=-\sum\limits_{i=1}^n\frac{d_i}{2m}\cdot\log_2\frac{d_i}{2m}.
\end{equation}

\end{theorem}
\begin{proof} By Definition \ref{def:structure entropy-T}, noting that for every $\alpha\in T$, let $g_{\alpha}=g(T_{\alpha})$ and $V_{\alpha}={\rm vol}(\alpha)$, we have:

\begin{eqnarray}
\mathcal{H}^T_g(G)&=&-\sum\limits_{\alpha\not=\lambda, \alpha\in\ T}
\frac{g_{\alpha}}{2m}\cdot \log_2\frac{V_{\alpha}}{V_{\alpha^{-}}} \nonumber\\
&=&-\sum\limits_{\alpha\not=\lambda, \alpha\in\ T}
\frac{g_{\alpha}}{2m}\cdot \log_2V_{\alpha} + \sum\limits_{\alpha\not=\lambda, \alpha\in\ T}
\frac{g_{\alpha}}{2m}\cdot \log_2V_{\alpha^{-}} \nonumber\\
&=&-\sum\limits_{\alpha\not=\lambda, \alpha\in\ T}
\frac{g_{\alpha}}{2m}\cdot \log_2V_{\alpha} + (\sum\limits_{\alpha\in T,\ \text{non-leaf}}
\frac{g_{\alpha}}{2m}\cdot \log_2V_{\alpha}+\log_2(2m)), \ \text{by the additivity of $g$} \nonumber\\
&=&-\sum\limits_{i=1}^n\frac{d_i}{2m}\cdot\log_2\frac{d_i}{2m}.
\end{eqnarray}

The theorem follows.
\end{proof}

\begin{definition}\label{def:length of random accessible vertex} (The length of the vertex accessible from random walk with stationary distribution)
For a connected, simple graph $G=(V,E)$ of $n$ vertices and $m$ edges. Let $g$ be the volume module function of $G$ defined as: for any set $X$ of vertices $V$, $g(X)$ is the volume of $X$.
Suppose that $T$ is a coding tree of $G$. Then:

\begin{equation}\label{eqn:Length-G-T}
H^T_g(G)=-\sum\limits_{\alpha\in T, \alpha\not=\lambda}\frac{V_{\alpha}}{2m}\cdot\log_2\frac{V_{\alpha}}{V_{\alpha^{-}}},
\end{equation}
where $V_{\beta}$ is the volume of $T_{\beta}$, $\alpha^{-}$ is the parent node of $\alpha$ in $T$.

\end{definition}

{\it Remarks: In this case, ${H}^T_g(G)$ is the amount of information required to describe the codeword in $T$ of the vertex that is accessible from random walk with stationary distribution in $G$, and is a lower bound of the average length of the codeword (in $T$) of the vertex that is accessible from random walk with stationary distribution in $G$.}

\begin{theorem} \label{thm:length of graph T} For any connected and simple graph $G=(V,E)$ with $n$ vertices and $m$ edges. For the module function $g(X)=\sum\limits_{x\in X}d_x$, where $d_x$ is the degree of $x$ in $G$, and for any coding tree $T$ of $G$,

\begin{eqnarray}
{H}^T_g(G)&=&-\sum\limits_{i=1}^n\frac{d_i}{2m}\cdot\log_2\frac{d_i}{2m} \nonumber\\
&=&\mathcal{H}^1(G),
\end{eqnarray}
where $d_i$ is the degree of vertex $i$ in $G$, $\mathcal{H}^1(G)$ is the one-dimensional structural entropy of $G$ \cite{LP2016a}.

\end{theorem}
\begin{proof} Note that for any non-leaf node $\alpha\in T$, $V_{\alpha}=\sum\limits_{\beta\in T, \beta^{-}=\alpha}V_{\beta}$, that is, $V_{\alpha}$ is an additive module function. The theorem follows from Theorem \ref{thm:additive function theorem}.
\end{proof}

Theorem \ref{thm:length of graph T} shows that

\begin{itemize}

\item The information to describe the codeword of a tree of the vertex that is accessible from random walk with stationary distribution in $G$ is independent of any coding tree $T$ of $G$, and
\item The {\it minimum average length}, written $L(G)$, of the codeword in a coding tree of the vertex that is accessible from random walk with stationary distribution is greater than or equal to (or lower bounded by) the one-dimensional structure entropy $\mathcal{H}^1(G)$ \cite{LP2016a}, or the Shannon entropy of the degree distribution of the graph. This means that

    \begin{equation}
    L(G)=\Omega (\log_2n),
    \end{equation}
where $n$ is the number of vertices in $G$.

This property is in sharp contrast to the structure entropy. In fact, there are many graphs $G$ such that the two-dimensional structure entropy $\mathcal{H}^2(G)=O(\log_2\log_2n)$, referred to \cite{LP2016a}.
\end{itemize}

The proof of Theorem \ref{thm:additive function theorem} also shows the reason why the structure entropy in Definitions \ref{def:structure entropy-T-O} and \ref{def:structure-entropy-O} depend on the coding trees of a graph. The reason is that, the cut module function $g$ in Definition \ref{def:structure entropy-T-O} fails to have the additivity, since for any two disjoint vertex sets $X$ and $Y$, if there are edges between $X$ and $Y$, then $g(X\cup Y)<g(X)+g(Y)$. This ensures that the structure entropy $\mathcal{H}^T(G)$ in Definition \ref{def:structure entropy-T-O} depends on the coding tree $T$ of $G$. For this reason, the structure entropy provides the foundation for a new direction of information theory with rich theory and remarkable applications in many areas.

\subsection{The relationship between Shannon entropy and the structure entropy}

Given a connected graph $G=(V,E)$, Theorem \ref{thm:length of graph T} implies that the information to describe the vertex that is accessible from random walk with stationary distribution cannot be reduced by any coding tree of the the graph, so that the average length of the codewords of the vertex that is accessible from random walk with stationary distribution must be lower bounded by the entropy of the degree distribution of the graph.

However, the structure entropy of a graph defined in Definition \ref{def:structure entropy-T-O} is determined by the coding tree of the graph, and hence the structure entropy is given in Definition \ref{def:structure-entropy-O}. We have seen in \cite{LP2016a}, that there are graphs $G$ whose two-dimensional structure entropy is $\mathcal{H}^2(G)=O(\log_2\log_2n)$. This means that coding trees play an essential role in controlling a network by reducing the uncertainty of the interactions in the network.

Theorems \ref{thm:additive function theorem} and \ref{thm:length of graph T}, together with the theory in \cite{LP2016a} imply that structure entropy of a graph involves one more measure of the graph, which is the module function of graphs.
This means that the variant of the structure entropy given in
Definitions \ref{def:structure entropy-T} and \ref{def:structural-entropy} is well-defined. In particular, using this variant of structure entropy, we are able to establish a new theory of structure entropy for directed graphs, in which case, the flow module function plays a role (project in progress).

The comparison between Shannon entropy and our structure entropy also suggests some interesting open questions. For example, Shannon entropy can be understood as the tight lower bound of the {\it length} of the binary expression of the item picked by the probabilistic distribution, and the tight lower bound for the number of bits required to guess the item chosen by the probabilistic distribution, and so on. Some of these measures such as the length of the binary expression is intuitive and concrete, and has geometric and physical meaning. However, the structure entropy has no such intuition. It is an open question to find a geometric or physical interpretation for the structure entropy in Definitions \ref{def:structure entropy-T-O} and \ref{def:structure-entropy-O}. Of course, the fundamental feature of structure entropy is that structure plays a role in information theory. This new feature leads to both information theoretical approach to graph theory and graphic approach to information theory, accompanying with the new notions of coding trees and module functions etc.

\section{Resistance of Networks}\label{sec:resistance}

In this section, we propose the notion of resistance and security index of a graph. The notions are built by using the one- and two-dimensional structure entropy introduced in \cite{LP2016a}.
We recall the one- and two-dimensional structure entropy \cite{LP2016a}.

\subsection{One- and two-dimensional structure entropy} \label{subsec:defPentropy}

According to Definition \ref{def:structure entropy-T-O}, the one-dimensional structure entropy of a graph $G$ has the following form:
Let $G=(V,E)$ be a connected graph with $n$ vertices and $m$ edges. For each vertex
$i\in \{1, 2, \cdots, n\}$, let $d_i$ be the degree of $i$ in $G$,
and let $p_i=d_i/2m$. Then the vector ${\bf p}=(p_1, p_2, \cdots,
p_n)$ is the stationary distribution of a random walk in $G$. By using this, we define the {\it one-dimensional structure entropy or
positioning entropy of $G$} by:

\begin{equation} \label{eqn:entropy_G}
\mathcal{H}^1(G)=H({\bf p})=H\left( \frac{d_1}{2m},\ldots,\frac{d_n}{2m}
\right)=-\sum\limits_{i=1}^n \frac{d_i}{2m} \cdot
\log_2\frac{d_i}{2m}.
\end{equation}

%\subsection{Two- and two-dimensional structure entropy}\label{subsec:defSentropy}

%To better understand the two-dimensional structure entropy, we introduce the following:

By Definition \ref{def:structure entropy-T-O}, we have:

\begin{definition} (Structure entropy of $G$ by a partition, \cite{LP2016a})\label{def:structureentropy-partition}
Given a connected graph $G=(V,E)$, suppose that $\mathcal{P}=\{X_1, X_2,
\cdots, X_L\}$ is a partition of $V$. We define the {\it structure
entropy of $G$ by $\mathcal{P}$} as follows:

\begin{eqnarray} \label{eqn:entropypartition}
\mathcal{H}^{\mathcal{P}}(G) &:=& \sum\limits_{j=1}^L\frac{V_j}{2m} \cdot
H\left( \frac{d_1^{(j)}}{V_j},\ldots,\frac{d_{n_j}^{(j)}}{V_j}
\right)-\sum\limits_{j=1}^L \frac{g_j}{2m} \log_2\frac{V_j}{2m} \nonumber \\
&=& -\sum\limits_{j=1}^L\frac{V_j}{2m} \sum\limits_{i=1}^{n_j}
\frac{d_i^{(j)}}{V_j} \log_2\frac{d_i^{(j)}}{V_j}-
\sum\limits_{j=1}^L \frac{g_j}{2m} \log_2\frac{V_j}{2m},
\end{eqnarray}
where $L$ is the number of modules in partition $\mathcal{P}$, $n_j$
is the number of vertices in module $X_j$, $d_i^{(j)}$ is the degree of the $i$-th vertex in $X_j$, $V_j$ is the volume of module $X_j$, and $g_j$
is the number of edges with exactly one endpoint in module $j$.
\end{definition}

According to the definition, $\mathcal{H}^{\mathcal{P}}(G)$ is the average number of bits required to determine the code $(i,j)$ of the vertex of the graph that is accessible from random walk with stationary distribution in $G$, where $i$ is the code of the vertex in its own module, and $j$ is the code of the module of the accessible vertex in the whole network $G$.

Now we turn to define the {\it two-dimensional structure entropy} of
$G$.

\begin{definition}(Two-dimensional structure entropy, \cite{LP2016a})\label{def:TwoD}
Given a connected graph $G$, define the structure entropy of $G$ by:

\begin{equation} \label{eqn:structureentropy}
\mathcal{H}^2(G)=\min\limits_{\mathcal{P}}\{ \mathcal{H}^{\mathcal{P}}(G)\},
\end{equation}
where $\mathcal{P}$ runs over all the partitions of $G$.
\end{definition}

 Clearly, the definition of $\mathcal{H}^2(G)$ in Definition \ref{def:TwoD} is the same as that in Definition \ref{def:structural-entropy} for $K=2$.

 For the one- and two-dimensional structure entropy, we will use some fundamental results from \cite{LP2016a}:

 \begin{theorem} (Lower bound of one-dimensional structure entropy of simple graphs)\label{thm:lowerbound_posi_simple}
Let $G=(V,E)$ be an undirected, connected, and simple graph with $m$ edges, i.e., $|E|=m$. Then:

 $$\mathcal{H}^1(G)\geq \frac{1}{2}\left(\log_2 m-1\right).$$
\end{theorem}

\begin{theorem} (Lower bound of one-dimensional structure entropy of graphs of balanced weights) \label{thm:lowerbound_posi_weighted}
Let $G=(V,E)$ be a connected graph with weight function $w$. Let
$m=|E|$ be the number of edges. If the ratio of maximum weight and
minimum weight is at most $m^\epsilon$, that is $\frac{\max_{e\in G}
\{w(e)\}}{\min_{e\in G} \{w(e)\}} \leq m^\epsilon$, for some
constant $\epsilon<1$, then:
$$\mathcal{H}^1(G)\geq\frac{1}{2}\left[(1-\epsilon)\log_2 m -1\right].$$
\end{theorem}

 Given a graph $G=(V,E)$, and a subset $S$ of $V$, the conductance of
$S$ in $G$ is given by

\begin{equation} \label{eqn:phisubset}
\Phi (S)=\frac{|E(S,\bar{S})|}{\min\{ {\rm vol}(S), {\rm vol
}(\bar{S})\}},
\end{equation}

\noindent where $E(S,\bar{S})$ is the set of edges with one
endpoint in $S$ and the other in the complement of $S$, i.e.
$\bar{S}$, and ${\rm vol}(X)$ is the sum of degrees $d_x$ for all
$x\in X$. The conductance of $G$ is defined to be the minimum of
$\Phi(S)$ over all subsets $S$'s, that is:

\begin{equation} \label{eqn:phiG}
\Phi (G)=\min\limits_{S\subset V}\{\Phi (S)\}.
\end{equation}

\begin{theorem} (Two-dimensional structure entropy principle)\label{thm:SEprinciple} For any graph $G$, the two-dimensional structure entropy of $G$ follows:

\begin{equation}
\mathcal{H}^2(G)\geq\Phi (G)\cdot \mathcal{H}^1(G),
\end{equation}
\noindent where $\Phi (G)$ is the conductance of $G$, and $\mathcal{H}^1(G)$ is the one-dimensional structure entropy of $G$.
\end{theorem}

 \begin{theorem} (Lower bounds of two-dimensional structure entropy of simple graphs)\label{thm:lowerbound_struc_simple}
Let $G=(V,E)$ be an undirected, connected and simple graph with number of
edges $|E|=m$. Then the two-dimensional structure entropy of $G$ satisfies

\begin{equation}\label{eqn:lowerboundsimple}
\mathcal{H}^2(G)=\Omega(\log_2\log_2 m).
\end{equation}
\end{theorem}

 [{\it Remark}: In \cite{LP2016a}, the authors first defined the one- and two-dimensional structure entropy and then extended to the high-dimensional cases. In that paper, we used the notion ``partitioning tree" in the definition of high dimensional structure entropy. Here we use the notion of coding tree of a graph. We would hope that this new notion is better for people to understand the structural information theory.]

The notion of structure entropy may have fundamental accompanying notions, for instance, noises, coding and decoding etc. The authors introduced the notion of resistance as an accompanying notion of structure entropy in \cite{LP2016a}. It is interesting that resistance is determined only by the one- and two-dimensional structure entropy of the graph.

%\subsection{Resistance of networks}

Given a network $G=(V,E)$, consider the following scenario:
Suppose that there is a virus which randomly spreads in $G$. Suppose that there is partition $\mathcal{P}$ of $G$ such that random walk in $G$ with stationary distribution easily goes to a small module $X$ of $\mathcal{P}$ after which it is not easy to escape from the module $X$. In this case, the spreading of the virus is restrained by the partition $\mathcal{P}$ of $G$. To characterise the scenario, we define the {\it resistance of $G$ given by partition $\mathcal{P}$}.

\begin{definition}(Resistance of a graph by a partition $\mathcal{P}$, Li and Pan \cite{LP2016a})\label{def:resistance-P} Let $G=(V,E)$ be a connected graph and $\mathcal{P}$ be a partition of $G$. The resistance of $G$ given by $\mathcal{P}$ is defined as follows:

\begin{equation}\label{eqn:resistance-G-P}
\mathcal{R}^{\mathcal{P}}(G)=
-\sum\limits_{j=1}^L \frac{V_j-g_j}{2m} \log_2\frac{V_j}{2m},
\end{equation}

\noindent where $V_j$ is the volume of the $j$-th module $X_j$ of $\mathcal{P}$, and $g_j$ is the number of edges from $X_j$ to the vertices outside $X_j$.

\end{definition}

We define the notion of {\it resistance of a graph $G$}.

\begin{definition}(Resistance of networks, Li and Pan \cite{LP2016a})\label{def:resistance} Let $G$ be a connected graph. We define the resistance of $G$ as follows:
\begin{equation}\label{eqn:resistance}
\mathcal{R}(G)=\max_{\mathcal{P}}\{\mathcal{R}^{\mathcal{P}}(G)\},
\end{equation}
where $\mathcal{P}$ runs over all the partitions of $G$.
\end{definition}

Intuitively, the resistance $\mathcal{R}(G)$ measures the force of $G$ to resist the cascading failure of virus attacks in $G$. As a matter of fact,
the authors and their coauthors have shown experimentally that the resistance of a network does measure the force of the network to resist cascading failures of virus attacks \cite{LHLP2016}, and that resistance maximization is the principle for defending the networks from virus attacks \cite{LZP2016}.

%\section{Resistance Law of Networks}

For the resistance of graph $G$ by $\mathcal{P}$, we have the following
{\it resistance principle of networks}.

\begin{theorem}(Resistance principle, Li and Pan, \cite{LP2016a})\label{thm:resistance-principle}
Let $G=(V,E)$ be a connected graph. Suppose that $\mathcal{P}$ is a partition of $V$ with the notations the same as that in the definitions of $\mathcal{H}^1(G)$ and $\mathcal{H}^{\mathcal{P}}(G)$.
 Then the positioning entropy of $G$, $\mathcal{H}^1(G)$, and the structure entropy of $G$ by given $\mathcal{P}$, i.e., $\mathcal{H}^{\mathcal{P}}(G)$, satisfy the following properties:

\begin{enumerate}

\item [(1)] (Additivity law of one-dimensional structure entropy) The positioning entropy of $G$ satisfies:
\begin{equation}\label{equL^U2}
\mathcal{H}^1(G)= -\sum\limits_{j=1}^L \frac{V_j}{2m} \sum\limits_{i=1}^{n_j}
\frac{d_i^{(j)}}{V_j} \log_2\frac{d_i^{(j)}}{V_j}-
\sum\limits_{j=1}^L \frac{V_j}{2m} \log_2\frac{V_j}{2m}.
\end{equation}

\item [(2)] (Local resistance law of networks)

\begin{eqnarray}
\mathcal{R}^{\mathcal{P}}(G)=
-\sum\limits_{j=1}^L \frac{V_j-g_j}{2m} \log_2\frac{V_j}{2m}=\mathcal{H}^1(G)-\mathcal{H}^{\mathcal{P}}(G)
\end{eqnarray}

\item [(3)] Assume that for each $j$, $V_j\leq m$, for $m=|E|$. Then

\begin{eqnarray}\label{eqn:PS}
\mathcal{R}^{\mathcal{P}}(G)=
-\sum\limits_{j=1}^L(1-\Phi(X_j)) \frac{V_j}{2m}
\log_2\frac{V_j}{2m}=\mathcal{H}^1(G)-\mathcal{H}^{\mathcal{P}}(G)
\end{eqnarray}

\noindent where $\Phi(X_j)$ is the conductance of $X_j$ in $G$.

\end{enumerate}
\end{theorem}

By the local resistance law in Theorem \ref{thm:resistance-principle}, we have:

\begin{theorem}(Global resistance law of networks, Li and Pan \cite{LP2016a})\label{thm:resistance} Let $G$ be a connected graph. Then, we have

\begin{equation}\label{eqn:resistanceprinciple}
\mathcal{R}(G)=\mathcal{H}^1(G)-\mathcal{H}^2(G).
\end{equation}
\end{theorem}
\begin{proof} By Theorem \ref{thm:resistance-principle}.
\end{proof}

According to Theorem \ref{thm:resistance}, we define the security index of a graph $G$ to be the normalised resistance of $G$. That is,

\begin{definition}(Security index of a graph)\label{def:securityindex} Let $G$ be a connected graph.
 We define the {\it security index of graph $G$} as follows:

\begin{equation}\label{eqn:resistanceratio}
\theta (G)=\frac{\mathcal{R}(G)}{\mathcal{H}^1(G)}.
\end{equation}
\end{definition}

 By the definition in Equation (\ref{eqn:resistance}) and the resistance principle of networks by partitions, we have that the resistance of a connected graph $G$ satisfies:

According to Theorem \ref{thm:resistance}, for a connected graph $G$, we have:

$$\mathcal{R}(G)=\mathcal{H}^1(G)-\mathcal{H}^2(G).$$

Notice that $\mathcal{H}^2(G)\leq\mathcal{H}^1(G)$.

By the definition of security index in Equation (\ref{eqn:resistanceratio}) and by the result in Equation (\ref{eqn:resistanceprinciple}), we have that the security index of a connected graph $G$ satisfies the following:

\begin{equation}
\theta (G)=1-\frac{\mathcal{H}^2(G)}{\mathcal{H}^1(G)}.
\end{equation}

Based on the security index, we introduce the following:

\begin{definition}\label{def:security-graph} (Resistor graph) Let $G=(V,E)$ be a connected graph of $n$ vertices and $m$ edges. Let $\theta$ be a number in $(0,1)$.
We say that $G$ is an $(n,\theta)$-resistor graph, if:

\begin{equation}
\theta (G)\geq\theta.
\end{equation}

\end{definition}

By Theorem \ref{thm:SEprinciple}, for any expander graph $G$, the conductance $\Phi (G)$ is a large constant $\alpha$, therefore, the security index of $G$ is $\theta (G)<1-\alpha$ for a large constant $\alpha$. This means that expanders are not good resistor graphs.

\section{Basic Theorems for Classic Structures}\label{sec:basic}

Trees and grid graphs perhaps are the most natural and most frequently used data structures. The authors \cite{LP2016a} have established some lower and upper bounds of the $K$-dimensional structure entropy of the graphs. Here we prove the basic
 theorems of the resistances
and security indices of the classical data structures.

\subsection{Resistance and security index of trees}

In this subsection, we consider the resistance and security indices of
complete binary trees. Similar results can be generalized easily to
any trees with constant bounded degrees. A complete binary tree is a
tree whose non-leaf nodes has exactly two children and every leaf
node has the same depth (In this section, for notational simplicity,
we define the depth of a node to be the number of nodes on the
unique path from this node to the root). So the complete binary tree
of depth $H$ has exactly $2^H-1$ nodes.

\begin{theorem} \label{thm:resis_tree}
Let $T$ be a complete binary tree of depth $H$ and thus of size
$n=2^H-1$. Then:

\begin{enumerate}
\item[(1)] The resistance of $T$ is $\mathcal{R}(T) \geq \log_2 n-\log_2\log_2 n -5-o(1)=(1-o(1))\cdot\log_2n$, and
\item[(2)] The security index of $T$ is $\theta(T) \geq 1-\frac{\log_2\log_2 n}{\log_2 n}-O\left(\frac{1}{\log_2 n}\right)=1-o(1)$.
\end{enumerate}

\end{theorem}

\begin{proof}
We will prove that
\begin{enumerate}
\item[(i)] $\mathcal{H}^1(T) \geq \log_2 n -1$, and
\item[(ii)] $\mathcal{H}^2(T) \leq \log_2\log_2 n +4+o(1)$.
\end{enumerate}
Then Theorem \ref{thm:resis_tree} follows immediately.

To calculate $\mathcal{H}^1(T)$, note that in $T$, there are $2^H-1$
nodes, and one of them is the root of degree $2$, $2^{H-1}$ of them
are leaves of degree $1$, and $2^{H-1}-2$ of them are intermediate
nodes of degree $3$. The total volume of $T$ is thus $2^{H+1}-4$. So
\begin{eqnarray*}
\mathcal{H}^1(T) &=& -\frac{2}{2^{H+1}-4}\log_2\frac{2}{2^{H+1}-4}
-(2^{H-1}-2)\cdot\frac{3}{2^{H+1}-4}\log_2\frac{3}{2^{H+1}-4}\\
&& -2^{H-1}\cdot\frac{1}{2^{H+1}-4}\log_2\frac{1}{2^{H+1}-4}\\
&=&
\frac{1}{2^H-2}\log_2(2^H-2)+\left[\frac{3(2^{H-2}-1)+2^{H-2}}{2^H-2}\right]\log_2(2^{H+1}-4)\\
&&-\frac{3(2^{H-2}-1)}{2^H-2}\log_2
3\\
&=&
\log_2(2^H-2)+\frac{2^H-3}{2^H-2}-\frac{3(2^H-4)}{4(2^H-2)}\log_2
3\\
&\geq& H-1\\
&\geq& \log_2 n -1.
\end{eqnarray*}

To prove $\mathcal{H}^2(T) \leq \log_2\log_2 n +4+o(1)$, it suffices
to define a partition $\mathcal{P}$ of the nodes in $T$ such that
$\mathcal{H}^\mathcal{P}(T)\leq \log_2\log_2 n +4+o(1)$. We define
$\mathcal{P}$ as follows. Let $1\leq k\leq H$ be an integer. We
partition every subtree whose root is a node of depth $H-k+1$ as a
module and the remaining part consisting of all the nodes of depth
at most $H-k$ as a module. Now we have $2^{H-k}$ complete binary
subtrees, each of which, denoted by $T_j$, $j=1,2,\ldots,2^{H-k}$,
has a size $2^k-1$ and another complete binary subtree, denoted by
$T'$, which has a size $2^{H-k}-1$. A simple calculation indicates
that for each $T_j$, its volume $\vol(T_j)=2^{k+1}-3$, and the
volume of $T'$ is $\vol(T')=3\cdot 2^{H-k}-4$.

For each $T_j$, we have
\begin{eqnarray*}
-\sum\limits_{v\in T_j} \frac{d_v}{2m} \log_2\frac{d_v}{\vol(T_j)}
&=&
-(2^{k-1}-1)\cdot\frac{3}{2m}\log_2\frac{3}{2^{k+1}-3}-2^{k-1}\cdot\frac{1}{2m}\log_2\frac{1}{2^{k+1}-3}\\
&\leq& \frac{1}{2m}\left[ (2^{k-1}-1)\cdot 3(k+1)+2^{k-1}(k+1)
\right]\\
&\leq& \frac{2^{k+1}}{2m}(k+1).
\end{eqnarray*}
So
\begin{eqnarray*}
-\sum\limits_{j=1}^{2^{H-k}} \frac{\vol(T_j)}{2m} \sum\limits_{v\in
T_j} \frac{d_v}{\vol(T_j)}\log_2\frac{d_v}{\vol(T_j)} &=&
-\sum\limits_{j=1}^{2^{H-k}} \sum\limits_{v\in T_j}
\frac{d_v}{\vol(T_j)}\log_2\frac{d_v}{\vol(T_j)}\\
&\leq& 2^{H-k} \cdot \frac{2^{k+1}}{2m}(k+1).
\end{eqnarray*}

Note that each $T_j$ has exactly one global edge connecting to $T'$.
So the number of global edges for each $T_j$ is $g_j=1$. We have
\begin{eqnarray*}
-\sum\limits_{j=1}^{2^{H-k}} \frac{g_j}{2m}\log_2
\frac{\vol(T_j)}{2m} &=& -2^{H-k}\cdot
\frac{1}{2m}\log_2\frac{2^{k+1}-3}{2m}\\
&=& \frac{2^{H-k}}{2m}\cdot \left[\log_2 2m -(k+1)
+O\left(\frac{1}{2^k}\right)\right].
\end{eqnarray*}

Then consider the subtree $T'$. Note that all the nodes in $T'$
except for the root of $T$ which has degree $2$, have degree $3$. So
\begin{eqnarray*}
-\sum\limits_{v\in T'} \frac{d_v}{2m} \log_2\frac{d_v}{\vol(T')} &=&
-(2^{H-k}-2)\cdot \frac{3}{2m}\log_2\frac{3}{3\cdot 2^{H-k}-4} -\frac{2}{2m}\log_2\frac{2}{3\cdot 2^{H-k}-4}\\
&\leq& \frac{2^{H-k}}{2m}\cdot 3(H-k).
\end{eqnarray*}

Note that $T'$ has $g_{T'}=2^{H-k}$ global edges, each of which
joins a subtree $T_j$. We have
\begin{eqnarray*}
-\frac{g_{T'}}{2m}\log_2\frac{\vol(T')}{2m} &=&
-\frac{2^{H-k}}{2m}\log_2\frac{3\cdot 2^{H-k}-4}{2m}\\
&=& \frac{2^{H-k}}{2m}\cdot \left[ \log_2 2m -(H-k)
+O\left(\frac{1}{2^{H-k}}\right) \right].
\end{eqnarray*}

So in all, noting that $\log_2 2m=\log_2(2^{H+1}-4)\leq H+1$, the
structure entropy of $T$ by partition $\mathcal{P}$ is
\begin{eqnarray*}
\mathcal{H}^\mathcal{P}(T) &=& -\sum\limits_{j=1}^{2^{H-k}}
\frac{\vol(T_j)}{2m} \sum\limits_{v\in T_j}
\frac{d_v}{\vol(T_j)}\log_2\frac{d_v}{\vol(T_j)}
-\sum\limits_{j=1}^{2^{H-k}} \frac{g_j}{2m}\log_2
\frac{\vol(T_j)}{2m}\\
&&-\sum\limits_{v\in T'} \frac{d_v}{2m} \log_2\frac{d_v}{\vol(T')}
-\frac{g_{T'}}{2m}\log_2\frac{\vol(T')}{2m}\\
&\leq& \frac{2^{H-k}}{2m}\cdot 2^{k+1}(k+1) +\frac{2^{H-k}}{2m}\cdot
\left[\log_2
2m -(k+1) +O\left(\frac{1}{2^k}\right)\right]\\
&& +\frac{2^{H-k}}{2m}\cdot 3(H-k) +\frac{2^{H-k}}{2m}\cdot \left[
\log_2 2m -(H-k) +O\left(\frac{1}{2^{H-k}}\right) \right]\\
&\leq& \frac{2^{H-k}}{2^{H+1}-4}\cdot \left[
(2^{k+1}+1)(k+1)+4(H-k)+O\left(\frac{1}{2^k}\right)+O\left(\frac{1}{2^{H-k}}\right) \right]\\
&\leq& (k+1)+\frac{4(H-k)}{2^{k+1}}+O\left(
\frac{k+1}{2^k}+\frac{k+1}{2^{H-k}}+\frac{H-k}{2^H} \right).
\end{eqnarray*}

When we choose $k+1=\lceil\log_2 H\rceil$, the above value is at
most $\lceil\log_2 H\rceil+4+o(1)$, which is $\log_2\log_2 n
+4+o(1)$. Theorem \ref{thm:resis_tree} follows.
\end{proof}

\subsection{Resistance and security index of grid graphs}

In this subsection, we consider the resistance and the security indices
of grid graphs. An $n\times n$ grid $G=(V,E)$ is a graph defined on the
vertex set $V=\{v_{i,j}:i,j\in\mathbb{Z^+}, 1\leq i,j\leq n\}$ and the
edge set $E=\{(v_{i,j},v_{i,j'}):|j-j'|=1\} \bigcup
\{(v_{i,j},v_{i',j}):|i-i'|=1\}$.

\begin{theorem} \label{thm:resis_grid}
Let $G=(V,E)$ be an $n\times n$ grid graph. Then the resistance and
the security index of $G$ satisfies:

\begin{enumerate}
\item[(1)] The resistance of $G$ is $\mathcal{R}(G) \geq \log_2[n(n-1)] -2\log_2\log_2 n -O(1))$, and
\item[(2)] The security index of $G$ is $\rho(G) \geq 1-\frac{2\log_2\log_2 n}{\log_2[n(n-1)]}-O\left(\frac{1}{\log_2 n}\right)$.
\end{enumerate}

\end{theorem}

\begin{proof}
We will prove that
\begin{enumerate}
\item[(i)] $\mathcal{H}^1(G) \geq \log_2[n(n-1)]$, and
\item[(ii)] $\mathcal{H}^2(G) \leq 2\log_2\log_2 n +O(1)$.
\end{enumerate}
Then Theorem \ref{thm:resis_grid} follows.

To calculate $\mathcal{H}^1(G)$, note that in a $n\times n$ grid,
there are four vertices (corners) of degree $2$, $4(n-2)$ vertices (sides)
of degree $3$ and $(n-2)^2$ vertices of degree $4$. The total volume of
$G$ is thus $4n(n-1)$. So
\begin{eqnarray*}
\mathcal{H}^1(G) &=& -4\cdot\frac{2}{4n(n-1)}\log_2\frac{2}{4n(n-1)}
-4(n-2)\frac{3}{4n(n-1)}\log_2\frac{3}{4n(n-1)}\\
&& -(n-2)^2\frac{4}{4n(n-1)}\log_2\frac{4}{4n(n-1)}\\
&=& \left[
\frac{2}{n(n-1)}+\frac{3(n-2)}{n(n-1)}+\frac{(n-2)^2}{n(n-1)}
\right]\cdot\log_2[n(n-1)]\\
&& +\frac{3(n-2)(2-\log_2 3)+2}{n(n-1)}\\
&=& \log_2[n(n-1)]+\frac{3(n-2)(2-\log_2 3)+2}{n(n-1)}\\
&\geq& \log_2[n(n-1)].
\end{eqnarray*}

To prove $\mathcal{H}^2(G) \leq 2\log_2\log_2 n +O(1)$, similarly to
the proof of Theorem \ref{thm:resis_tree}, we find a partition
$\mathcal{P}$ for the vertices in $G$ to witness the upper bound. We
divide $G$ into sub-grids of size $k\times k$. For notational
simplicity, assume that $n$ can be divided by $k$. So we have
exactly $\left(\frac{n}{k}\right)^2$ such sub-grids. For each
sub-grid, denoted by $G_j$, let $d_i^{(j)}$ denote the degree of the
$i$-th vertex, which is $4$ for most vertices, $3$ for border vertices and
$2$ for corner vertices of $G$. By the extremum property of the entropy
function $H(\cdot)$, the positioning entropy within $G_j$ satisfies
$$
H\left(\frac{d_1^{(j)}}{\vol(G_j)},\cdots,\frac{d_{k^2}^{(j)}}{\vol(G_j)}\right)
\leq \log_2 k^2 = 2\log_2 k.
$$
So
$$
\sum\limits_j \frac{\vol(G_j)}{2m}\cdot
H\left(\frac{d_1^{(j)}}{\vol(G_j)},\cdots,\frac{d_{k^2}^{(j)}}{\vol(G_j)}\right)
\leq 2\log_2 k.
$$
Since the total number of global edges is
$$\sum\limits_j g_j=2n\left( \frac{n}{k}-1 \right),$$ and noting that $m=2n(n-1)$, we have
$$-\sum\limits_j \frac{g_j}{2m}\log_2 \frac{\vol(G_j)}{2m}
\leq \left(\sum\limits_j g_j\right)\cdot\frac{1}{2m}\log_2 2m \leq
\frac{n-k}{2k(n-1)}\cdot(2\log_2 n +2) \leq \frac{\log_2 n +1}{k}.$$

So in all, we have that the structure entropy of $G$ by partition
$\mathcal{P}$ is

\begin{eqnarray*}
\mathcal{H}^{\mathcal{P}}(G) &=& \sum\limits_j
\frac{\vol(G_j)}{2m}\cdot
H\left(\frac{d_1^{(j)}}{\vol(G_j)},\cdots,\frac{d_{k^2}^{(j)}}{\vol(G_j)}\right)-\sum\limits_j
\frac{g_j}{2m}\log_2 \frac{\vol(G_j)}{2m}\\
&\leq& 2\log_2 k +\frac{\log_2 n +1}{k}.
\end{eqnarray*}

Let $k=\Theta(\log_2 n)$, then $\mathcal{H}^{\mathcal{P}}(G)\leq
2\log_2 \log_2 n +O(1)$. Theorem \ref{thm:resis_grid} follows.
\end{proof}

Theorems \ref{thm:resis_tree} and \ref{thm:resis_grid} show that the classical graphs such as trees and grid graphs can be used as the basic module of secure networks. This is a surprising result, since it means that secure networks may be constructed by using the basic structures.

\section{Resistance of Bounded Degree Graphs}\label{sec:bounded degree graph}

In this section, we give a lower bound for the resistance of regular
graphs.

\begin{theorem}\label{thm:lowerbound-R}
Let $G=(V,E)$ be a simple, connected graph of bounded degree $d$ for
some constant $d$. Then
\begin{eqnarray}
\mathcal{R}(G)\geq\left(\frac{2}{d}-o(1)\right)\cdot\log_2 n.
\end{eqnarray}
\end{theorem}

\begin{proof}
We only have to show that there is a partition $\mathcal{P}=V_1\cup
V_2\cup\cdots\cup V_L$, such that
\begin{equation} \label{eqn:H1HPdifference}
\mathcal{R}(G)\geq\mathcal{H}^1(G)-\mathcal{H}^\mathcal{P}(G)
=-\sum\limits_{j=1}^L(1-\Phi(V_j))\cdot\frac{\vol(V_j)}{\vol(G)}\log_2\frac{\vol(V_j)}{\vol(G)}
\geq\left(\frac{2}{d}-o(1)\right)\cdot\log_2 n.
\end{equation}

Since $G$ is connected, consider a spanning tree, denoted by $T$, of
$G$. Since $G$ has bounded degree, so is $T$. Pick an arbitrary
vertex $r$ of $T$ as the root. Then the depth of every other vertex
$v$ is the length of the unique path from $r$ to $v$, and every
vertex on this path other than $v$ is called an ancestor of $v$. We
say that the $k$-th ancestor of $v$ is the one which has distance
$k$ from $v$.

We define a partition of vertices in $T$ recursively in the
following way. Let $l=\lfloor\log_d\log_2 n\rfloor-1$ and denote
$T_0=T$. For $i=0,1,2,\ldots$, find the deepest vertex, denoted by
$v_i$, of $T_i$ (break ties arbitrarily), and denote by $v_{i+1}$
the ancestor of $v_i$ which has distance $l$ from $v_i$. Take the
subtree rooted by $v_{i+1}$, denoted by $V_{i+1}$, as a module, and
then delete $T_{i+1}$ from $T_i$. This procedure will not end until
$T_i$ is empty. Suppose that the last module is $V_L$. Then
$\mathcal{P}\triangleq V_1\cup V_2\cup\cdots\cup V_L$ is a partition
of $V$. Next, we show that for this partition $\mathcal{P}$,
Inequality (\ref{eqn:H1HPdifference}) is satisfied.

Note that each $T_i$ has bounded degree $d$, and so the size of each
$T_i$ is at most $d^{l+1}\leq\log_2 n$, and $L\geq n/\log_2 n$.
Except for the last module $V_L$, the size of each $V_i$ is certainly at
least $l$. For each $V_i$, since it is connected in the spanning
tree $T$ of $G$, it is also connected in $G$. So there are at least
$|V_i|-1$ edges in $G$ whose two endpoints are both in $V_i$, and
$\vol(V_i)\geq 2(|V_i|-1)+1$. Thus, for each $V_i$, we have
$$\Phi(V_i)\leq\frac{\vol(V_i)-2(|V_i|-1)}{\vol(V_i)}\leq\frac{d|V_i|-2(|V_i|-1)}{d|V_i|}=1-\frac{2}{d}+\frac{2}{d|V_i|}.$$
Therefore, for sufficiently large $n$ \footnote{In this paper, whenever we say a proposition holds for ``sufficiently large" values, we mean that there is some large enough value such that the proposition holds for all values larger than this one},
\begin{eqnarray*}
\mathcal{R}(G) &\geq& \mathcal{H}^1(G)-\mathcal{H}^\mathcal{P}(G)\\
&=&
-\sum\limits_{j=1}^L(1-\Phi(V_j))\cdot\frac{\vol(V_j)}{\vol(G)}\log_2\frac{\vol(V_j)}{\vol(G)}\\
&\geq&
-\sum\limits_{j=1}^L\left(\frac{2}{d}-\frac{2}{d|V_j|}\right)\cdot\frac{\vol(V_j)}{\vol(G)}\log_2\frac{\vol(V_j)}{\vol(G)}\\
&\geq&
-\sum\limits_{j=1}^{L-1}\left(\frac{2}{d}-\frac{2}{dl}\right)\cdot\frac{\vol(V_j)}{\vol(G)}\log_2\frac{\vol(V_j)}{\vol(G)}\\
&\geq& \frac{2}{d}\cdot\left(1-\frac{1}{l}\right)\cdot\frac{\vol(G)
-\vol(V_L)}{\vol(G)}\cdot\left(-\sum\limits_{j=1}^{L-1}\frac{\vol(V_j)}{\vol(G)-\vol(V_L)}\log_2\frac{\vol(V_j)}{\vol(G)-\vol(V_L)}\right)\\
&\geq& \frac{2}{d}\cdot\left(1-\frac{1}{l}\right)\cdot\frac{\vol(G)
-\vol(V_L)}{\vol(G)}\cdot\log_2\frac{\vol(G)-\vol(V_L)}{d\log_2 n}\\
&\geq& \frac{2}{d}\cdot\left(1-\frac{1}{l}\right)\cdot\frac{\vol(G)
-d\log_2 n}{\vol(G)}\cdot\log_2\frac{\vol(G)-d\log_2 n}{d\log_2 n}\\
&=& \left(\frac{2}{d}-o(1)\right)\cdot\log_2 n.
\end{eqnarray*}
This completes the proof of the theorem.
\end{proof}

\section{Resistance of Complete Graphs}\label{sec:complete graph}

As mentioned before, Theorem \ref{thm:SEprinciple} indicates that expander graphs are not good resistor graphs.

In this section, we analyze the resistance of the ``most expanding"
graphs, i.e., the complete graphs. We show that the resistance of a complete
graph is as low as a constant $O(1)$. First, we answer the following
question: When a partition $\mathcal{P}$ is given on a graph, to
achieve a larger resistance (or equivalently, a smaller two-dimensional structural
information) from $\mathcal{P}$, how to split or merge
the modules in $\mathcal{P}$. For two subsets of vertices $X$ and
$Y$, denote by $e(X,Y)$ the number of edges whose one endpoint is in
$X$ and the other in $Y$. The following lemma gives the
criteria.

\begin{lemma} \label{lem:merge_split} (Merging-Splitting Lemma)
Let $G=(V,E)$ be a regular graph. Let $\mathcal{P}_1=X_1\cup
X_2\cup\cdots\cup X_L$ and $\mathcal{P}_2=Y_1\cup Y_2\cup
X_2\cup\cdots\cup X_L$ be two partitions of $V$ whose only
difference is the module $X_1$ in $\mathcal{P}_1$ being split into
two modules $Y_1\cup Y_2$ in $\mathcal{P}_2$. Then
$\mathcal{H}^{\mathcal{P}_2}(G)\geq\mathcal{H}^{\mathcal{P}_1}(G)$
if and only if
$$e(Y_1,Y_2)\cdot\log_2\frac{n}{|X_1|} \geq e(Y_1,Y_1)\cdot\log_2\frac{|X_1|}{|Y_1|} + e(Y_2,Y_2)\cdot\log_2\frac{|X_1|}{|Y_2|},$$
and
$\mathcal{H}^{\mathcal{P}_2}(G)\leq\mathcal{H}^{\mathcal{P}_1}(G)$
if and only if
$$e(Y_1,Y_2)\cdot\log_2\frac{n}{|X_1|} \leq e(Y_1,Y_1)\cdot\log_2\frac{|X_1|}{|Y_1|} + e(Y_2,Y_2)\cdot\log_2\frac{|X_1|}{|Y_2|}.$$
\end{lemma}

\begin{proof}
The proof is straightforward. For any non-empty subset of vertices
$X$, let $H(X)$ denote the entropy of the degree distribution of
vertices in $X$. By Definition \ref{def:structureentropy-partition},
\begin{eqnarray*}
\mathcal{H}^{\mathcal{P}_2}(G)-\mathcal{H}^{\mathcal{P}_1}(G) &=&
\left( \frac{\vol(Y_1)}{\vol(G)}\cdot H(Y_1) +
\frac{\vol(Y_2)}{\vol(G)}\cdot H(Y_2) -
\frac{\vol(X_1)}{\vol(G)}\cdot H(X_1) \right)\\
&& - \left(
\Phi(Y_1)\cdot\frac{\vol(Y_1)}{\vol(G)}\log_2\frac{\vol(Y_1)}{\vol(G)}
+
\Phi(Y_2)\cdot\frac{\vol(Y_2)}{\vol(G)}\log_2\frac{\vol(Y_2)}{\vol(G)} \right.\\
&& \left.-
\Phi(X_1)\cdot\frac{\vol(X_1)}{\vol(G)}\log_2\frac{\vol(X_1)}{\vol(G)} \right)\\
&=& \left(\frac{\vol(Y_1)}{\vol(G)}\cdot\log_2|Y_1| +
\frac{\vol(Y_2)}{\vol(G)}\cdot\log_2|Y_2| -
\frac{\vol(Y_1)+\vol(Y_2)}{\vol(G)}\cdot\log_2|X_1|\right)\\
&& - \left(
\frac{e(Y_1,\overline{Y}_1)}{\vol(G)}\log_2\frac{\vol(Y_1)}{\vol(G)}
+
\frac{e(Y_2,\overline{Y}_2)}{\vol(G)}\log_2\frac{\vol(Y_2)}{\vol(G)} \right.\\
&& \left.-
\frac{e(X_1,\overline{X}_1)}{\vol(G)}\log_2\frac{\vol(X_1)}{\vol(G)}
\right)\\
&& \text{(Note that
$e(X_1,\overline{X}_1)=e(Y_1,\overline{Y}_1)+e(Y_2,\overline{Y}_2)-2e(Y_1,Y_2)$)}\\
&=& \left( \frac{\vol(Y_1)}{\vol(G)}\cdot\log_2\frac{|Y_1|}{|X_1|} +
\frac{\vol(Y_2)}{\vol(G)}\cdot\log_2\frac{|Y_2|}{|X_1|} \right)\\
&& - \left(
\frac{e(Y_1,\overline{Y}_1)}{\vol(G)}\cdot\log_2\frac{\vol(Y_1)}{\vol(X_1)}
+
\frac{e(Y_2,\overline{Y}_2)}{\vol(G)}\cdot\log_2\frac{\vol(Y_2)}{\vol(X_1)}
\right.\\
&& \left. +
\frac{2e(Y_1,Y_2)}{\vol(G)}\cdot\log_2\frac{\vol(X_1)}{\vol(G)}
\right)\\ && \text{(Note that
$G$ is regular)}\\
&=&
\frac{\vol(Y_1)-e(Y_1,\overline{Y}_1)}{\vol(G)}\cdot\log_2\frac{|Y_1|}{|X_1|}
+
\frac{\vol(Y_2)-e(Y_2,\overline{Y}_2)}{\vol(G)}\cdot\log_2\frac{|Y_2|}{|X_1|}\\
&& - \frac{2e(Y_1,Y_2)}{\vol(G)}\cdot\log_2\frac{\vol(X_1)}{\vol(G)}\\
&=& \frac{2e(Y_1,Y_1)}{\vol(G)}\cdot\log_2\frac{|Y_1|}{|X_1|} +
\frac{2e(Y_2,Y_2)}{\vol(G)}\cdot\log_2\frac{|Y_2|}{|X_1|} -
\frac{2e(Y_1,Y_2)}{\vol(G)}\cdot\log_2\frac{|X_1|}{n}.
\end{eqnarray*}
So
$\mathcal{H}^{\mathcal{P}_2}(G)\geq\mathcal{H}^{\mathcal{P}_1}(G)$
if and only if
$$e(Y_1,Y_2)\cdot\log_2\frac{n}{|X_1|} \geq e(Y_1,Y_1)\cdot\log_2\frac{|X_1|}{|Y_1|} + e(Y_2,Y_2)\cdot\log_2\frac{|X_1|}{|Y_2|},$$
and
$\mathcal{H}^{\mathcal{P}_2}(G)\leq\mathcal{H}^{\mathcal{P}_1}(G)$
if and only if
$$e(Y_1,Y_2)\cdot\log_2\frac{n}{|X_1|} \leq e(Y_1,Y_1)\cdot\log_2\frac{|X_1|}{|Y_1|} + e(Y_2,Y_2)\cdot\log_2\frac{|X_1|}{|Y_2|}.$$
The lemma follows.
\end{proof}

By Lemma \ref{lem:merge_split}, we know that to reduce the structure
information, a large module (large $|X_1|$) tends to split into
pieces (in the case that
$\mathcal{H}^{\mathcal{P}_2}(G)<\mathcal{H}^{\mathcal{P}_1}(G)$),
while small modules (small $|Y_1|$ and $|Y_2|$) tend to merge into
big ones (in the case that
$\mathcal{H}^{\mathcal{P}_2}(G)>\mathcal{H}^{\mathcal{P}_1}(G)$).

For complete graphs, we have the following theorem.

\begin{theorem} \label{thm:complete_graph}
Let $G$ be a complete graph with $n$ vertices. Then
\begin{eqnarray}
\mathcal{R}(G)=O(1).
\end{eqnarray}
\end{theorem}

\begin{proof}
In the complete graph $G$ of size $n$, since each vertex has degree
$n-1$, a subset of vertices of size $x$ has volume $(n-1)x$ and
conductance $(n-x)/(n-1)$, $\vol(G)=n(n-1)$.

Suppose that $\mathcal{P}=V_1\cup V_2\cup\cdots\cup V_L$ be the
partition of $V$ such that
$\mathcal{H}^2(G)=\mathcal{H}^\mathcal{P}(G)$. Let $n_j=|V_j|$ for
each $j\in [L]$. Then
\begin{eqnarray*}
\mathcal{R}(G) &=& -\sum\limits_{j=1}^L
(1-\Phi(V_j))\cdot\frac{\vol(V_j)}{\vol(G)}\log_2\frac{\vol(V_j)}{\vol(G)}\\
&=& -\sum\limits_{j=1}^L \left( 1-\frac{n-n_j}{n-1}
\right)\cdot\frac{n_j}{n}\log_2\frac{n_j}{n}\\
&=& -\sum\limits_{j=1}^L
\frac{(n_j-1)n_j}{(n-1)n}\log_2\frac{n_j}{n}.
\end{eqnarray*}
Next, we prove that when $\mathcal{P}$ makes
$\mathcal{H}^\mathcal{P}(G)$ minimized, the size of modules in
$\mathcal{P}$ should be the same \footnote{Since the size of a
module should be an integer, here we say that two modules has the
same size if the deficit is at most one. But for the notational
simplicity, we assume that $n$ can always be divided by the
parameters we suppose, and the error will be absorbed in the
notation $O(\cdot)$.}. Suppose that $x$ and $y$ are the sizes of two
modules and $x+y=a$ for some fixed $a$. We will show that when other
modules are fixed, averaging $x$ and $y$, that is $x=y=a/2$, or
$x=0$ while $y=a$, or $x=a$ while $y=0$ makes the structural
information (under this partition) minimized. This means that for
two modules in a partition, to reduce the structure information,
they tend to be have the same size, or merge into a single module.
This holds for every pair of modules, which implies that the module
sizes in the optimal partition are averaged.

Note that $$(n-1)n\cdot\mathcal{R}(G)=-\sum\limits_{j=1}^L
(n_j-1)n_j\cdot\log_2\frac{n_j}{n}.$$ We only have to show that the
function
$$f(x)\triangleq -x(x-1)\cdot\log_2\frac{x}{n}-(a-x)(a-x-1)\cdot\log_2\frac{a-x}{n}$$
achieves maximum at $x=a/2$, or $0$, or $a$, when $0\leq x\leq a$.
The first derivative and the second derivative of $f(x)$ satisfies
\begin{eqnarray*}
\ln 2\cdot f'(x) &=& -(2x-1)\cdot\ln x-[2(x-a)+1]\cdot\ln
(a-x)+2(2x-a)\cdot\ln n+(a-2x),\\
\ln 2\cdot f''(x) &=& -2\cdot\ln [x(a-x)]+\frac{a}{x(a-x)}+4\cdot\ln
n-6.
\end{eqnarray*}
Note that the function
$$g(x)=-2\cdot\ln x+\frac{a}{x}+4\cdot\ln
n-6$$ strictly decreases monotonically for $x>0$. There is at most
one root for $g(x)$, and consequently, there are at most two roots
for $f''(x)$, and thus, there are at most two inflection points for
$f'(x)$. Observing that $f'(a/2)=0$, $\lim_{x\rightarrow 0+}
f'(x)=-\infty$ and $\lim_{x\rightarrow a-} f'(x)=+\infty$, we know
that there are at most three maximal values and two minimal values
for $f(x)$ in the interval $[0,a]$ because of at most two inflection
points in this interval, and these three (possible) maximal points
take values at $x=0$ or $a/2$ or $a$. This means that $x=y=a/2$, or
$x=0$ while $y=a$, or $x=a$ while $y=0$ makes the structural
information minimized when other modules are fixed.

So from now on, we can suppose that $x$ is the size of each modules
in $\mathcal{P}$, and so $L=n/x$ (suppose that $n$ can be divided by
$x$). We have
$$(n-1)n\cdot\mathcal{R}(G)=-\frac{n}{x}(x-1)x\cdot\log_2\frac{x}{n}=-n(x-1)\cdot\log_2\frac{x}{n}.$$
Define function
$$h(x)\triangleq \ln 2\cdot (n-1)\cdot\mathcal{R}(G)=-(x-1)\cdot\ln\frac{x}{n}.$$
To compute the maximum value of $h(x)$, note that
\begin{eqnarray*}
h'(x) &=& \ln\frac{n}{x}+\frac{1}{x}-1,\\
h''(x) &=& -\frac{x+1}{x^2}<0.
\end{eqnarray*}
Thus, $h(x)$ takes the maximum value at $x=x_0$ where $x_0$ is the
unique root of $h'(x)$. That is,
$$\ln\frac{x_0}{n}=\frac{1}{x_0}-1.$$
Therefore,
$$h(x_0)=-(x_0-1)\cdot\ln\frac{x_0}{n}=x_0+\frac{1}{x_0}-2\leq n+\frac{1}{n}-2.$$
So
$$\mathcal{R}(G)\leq\frac{n+\frac{1}{n}-2}{\ln 2\cdot(n-1)}=\frac{1}{\ln 2}\cdot\frac{n-1}{n}<\log_2 e.$$
Adding the error caused by the deficit of module sizes, we have
$\mathcal{R}(G)=O(1).$ This completes the proof of the theorem.
\end{proof}

In the above proof, note that $h'(n/e)>0$, and when $n\geq 7$,
$h'(n/2)<0$, which means that $n/e<x_0<n/2$, and in the optimal
partition $\mathcal{P}$, the number of modules $L=2$ or $3$ while
each module has size $n/2$ or $n/3$ for $n\geq 7$. Theorem
\ref{thm:complete_graph} means that any partition $\mathcal{P}$ of
the complete graph saves only a constant bits of information.

Theorem
\ref{thm:complete_graph} indicates that the resistance of the complete graphs is $O(1)$, and the security index of the complete graphs is $O(\frac{1}{\log_2n})=o(1)$, where $n$ is the number of vertices of the graph, so that the complete graphs are far from resistor graphs.

The arguments above clearly indicate that

\begin{theorem} For arbitrarily small constant $\epsilon>0$, there is an $N$ such that for any $n\geq N$, the complete graph of $n$ vertices cannot be an $(n,\epsilon)$-resistor graph.

\end{theorem}
\begin{proof} By the proof of Theorem
\ref{thm:complete_graph}.
\end{proof}

\section{Resistance and Security Index of the Networks of the Security Model}\label{sec:security}

Li, Li, Pan and Zhang \cite{LLPZ2015} proposed the security model of networks.
Here we establish the theory of resistance and security index of the networks generated by the security model.

The model proceeds as
follows.

\begin{definition} \label{def:securitymodel} (Security model, \cite{LLPZ2015}) Given a {\it homophyly (or affinity) exponent
$a\geq 0$} and a natural number $d$,

\begin{enumerate}

\item [1)] Let $G_{n_0}$ be an initial graph of size $n_0$ such that each vertex
has a distinct {\it color} and called {\it seed}, where $n_0$ is an
arbitrary positive integer.

For each step $i>n_0$, let $G_{i-1}$ be the graph constructed at the
end of step $i-1$, and $p_i=1/(\log i)^a$.

\item [2)] At step $i$, we create a new vertex, $v$.

\item [3)] With probability $p_i$, $v$ chooses a new color, in which case,

\begin{enumerate}
\item [i)] we call $v$ a seed,
\item [ii)] (preferential attachment) create an edge $(v,u)$ where $u$ is
chosen with probability proportional to the degrees of vertices in
$G_{i-1}$, and
\item [iii)] (randomness) create $d-1$ edges $(v,u_j)$,
where each $u_j$ is chosen randomly and uniformly among all seed
vertices in $G_{i-1}$.
\end{enumerate}

\item [4)] Otherwise, then $v$ chooses an old color, in which case,

\begin{enumerate}
\item [i)] (randomness) $v$ chooses uniformly and randomly an old color as
its own color and
\item [ii)] (homophyly and preferential attachment)
create $d$ edges $(v,u_j)$, where $u_j$ is chosen with probability
proportional to the degrees of all vertices of the same color as that
of $v$ in $G_{i-1}$.
\end{enumerate}
\end{enumerate}
\end{definition}

We use $\mathcal{S}(n,a,d)$ to denote the model with affinity exponent $a$, average number of edges $d$ and number of vertices $n$.

The authors \cite{LP2016} have shown that

\begin{enumerate}

\item (Uniform threshold security theorem)
 Let $G$ be a graph constructed from
$\mathcal{S}(n,a,d)$ with $p_i=\log^{-a} i$ for homophyly exponent
$a>4$ and for $d\geq 4$. Let the uniform threshold
$\phi=O\left(\frac{1}{\log^b n}\right)$ for
$b=\frac{a}{2}-2-\epsilon$ for arbitrarily small $\epsilon>0$.

Then with probability $1-o(1)$ (over the construction of $G$), there
is no initial set of poly-logarithmic size which causes a cascading
failure set of non-negligible size. Precisely, we have that for any
constant $c>0$,
$$\Pr_{{}G\in_{\rm R}\mathcal{S}(n,a,d),\ G=(V,E)}\left[\forall
S\subseteq V,\ |S|=\lceil\log^c n\rceil,\ |{\rm
inf}_G^\phi(S)|=o(n)\right]=1-o(1),$$

\noindent where ${\rm inf}_G^{\phi}(S)$ is the infection set of $S$
in $G$ with uniform threshold $\phi$.

\item (Random threshold security theorem) Let $a>6$ be the homophyly
exponent, and $d\geq 4$. Suppose that $G$ is a graph generated from
$\mathcal{S}(n,a,d)$.

Then with probability $1-o(1)$ (over the construction of $G$), there
is no initial set of poly-logarithmic size which causes a cascading
failure set of non-negligible size. Formally, we have that for any
constant $c>0$,
$$\Pr_{{}G\in_{\rm R}\mathcal{S}(n,a, d), \ G=(V,E)}\left[\forall
S\subseteq V,|S|=\lceil\log^c n\rceil,|{\rm inf}_G^{\rm
R}(S)|=o(n)\right]=1-o(1).$$

\end{enumerate}

In the present paper, we establish the resistance and security indices of the network of the security model.

\begin{theorem}
(Resistance theorem of the networks of the security model)
\label{thm:resistance_security} Given affinity exponent $a>0$ and
natural number $d>1$, let $G=(V,E)$ be a network of the security
model with $n$ vertices, affinity exponent $a$ and average number of
edges $d$. Then, with probability $1-o(1)$,

\begin{enumerate}
\item the resistance of $G$ is $\mathcal{R}(G)=\Omega (\log n)$, and
\item the security index of $G$ is $\theta (G)=1-o(1)$.
\end{enumerate}
\end{theorem}

Theorem \ref{thm:resistance_security} ensures that for every affinity exponent $a>0$ and for every edge parameter $d$, for the networks of the security model with affinity $a$ and edge parameter $d$ and with sufficiently large number of vertices $n$, with probability $\approx 1$, the resistances of the networks are as large as $\Omega (\log n)$, and the security indices of the networks are close to $1$. Therefore the networks are secure against cascading failures of arbitrarily strategic virus attacks.

The proof of Theorem \ref{thm:resistance_security} consists of two parts, the first part is a lower bound of one-dimensional structure entropy of the networks, and the second part is the two-dimensional structure entropy of the networks.

For the first part, we use
Theorems \ref{thm:lowerbound_posi_simple} and \ref{thm:lowerbound_posi_weighted}.

Therefore, for simple or balanced graphs $G$, we have:

\begin{equation}\label{eqn:OneDstructureentropy}
\mathcal{H}^1(G)=\Omega (\log n).
\end{equation}

For the second part, we investigate the structure entropy of the networks given by the natural partition classified by colors.

Let $G=(V,E)\in\mathcal{S}(n,a,d)$ be a network of $n$ vertices
generated from our Security model. Every vertex is associated with a color, we say that the classification of the vertices by colors is the natural community structure of $G$.
In so doing, a natural community of $G$ is a maximal homochromatic set. We note that each natural community contains a seed, which is the first vertex specified in the community.
We use $\mathcal{N}$ to denote the natural community structure of $G$.

We approximate the two-dimensional structure entropy of $G$ by $\mathcal{H}^{\mathcal{N}}(G)$.

First, we introduce some notations and basic probabilistic tools.

For every $t$, we use $G_t$ to denote the graph
obtained at the end of time step $t$ of the construction of $G$, and
$C_t$ to denote the set of seed vertices of $G_t$.

For
an edge $e=(u,v)$, we call $e$ a {\it local edge}, if the two
endpoints $u$, $v$ share the same color, and a {\it global edge},
otherwise.

The probabilistic bounds used are referred to the appendix.

For estimating $\mathcal{H}^{\mathcal{N}}(G)$, we establish the following fundamental properties of the networks generated by the security model.

\begin{theorem} \label{thm:basicproperty}
(Fundamental theorem of the networks of the security model) Given $a\geq 0$ and $d\geq 2$, let $G=(V,E)$ be a
graph of $n$ vertices generated from $\mathcal{S}(n, a, d)$. Let
$T_1=\log^{a+1}n$ and $T_2=\frac{n}{\log^b n}$ for some positive
constant $b$. Then the following properties hold.

\begin{enumerate}

\item [(1)] With probability $1-o(1)$,
for all $t\geq T_1$, $\frac{t}{2\log^a t}\leq |C_t|\leq
\frac{2t}{\log^a t}$.

\item [(2)] When $a>0$, for each homochromatic set $S$, if $t>t_S\geq T_1$, then the expectation of its
size at time step $t$ is $\Theta(\log^{a+1}t-\log^{a+1}t_S)$, where
$t_S$ is the time step at which the seed vertex of $S$ is created.

\item [(3)] With probability $1-o(1)$, every
homochromatic set in $G$ has a size upper bounded by $4\log^{a+1}n$.

\item [(4)] For each homochromatic set $S$, if $t_S\geq T_2$, then for sufficiently large $n$
the number of global edges in $G$ associated to $S$, denoted by
$g_S$, satisfies that
\begin{enumerate}
\item[(i)] if $a>1$, then $E(g_S)\leq \frac{5}{2}(a+1)b^2(\log\log
n)^2$;
\item[(ii)] if $a=1$, then $E(g_S)\leq 8b^2(\log\log
n)^2$;
\item[(iii)] if $0<a<1$, then $E(g_S)\leq 5b^2 (\log\log
n)^2$.
\end{enumerate}

\end{enumerate}
\end{theorem}

The proof of Theorem \ref{thm:basicproperty} is referred to the appendix.

Then we turn to prove Theorem \ref{thm:resistance_security}.

\begin{proof} (Proof of Theorem \ref{thm:resistance_security})
According to Equation (\ref{eqn:OneDstructureentropy}), it
suffices to show that, when $a>0$, with probability $1-o(1)$, the
two-dimensional structure entropy of $G$ is $\mathcal{H}^2(G)=o(\log
n)$, which is negligible compared to its one-dimensional structure
entropy $\mathcal{H}^1(G)$. Thus, the resistance of $G$ is
approximately $\mathcal{H}^1(G)$, which is $\Omega(\log n)$, and the
security index of $G$ is $1-o(1)$.

To establish an upper bound for $\mathcal{H}^2(G)$, it suffices to give a
partition for the vertices in $G$ with
$\mathcal{H}^{\mathcal{P}}(G)=o(\log n)$. Let $\mathcal{N}$ be the
natural partition given by the homochromatic sets.

Recall Equation (\ref{eqn:entropypartition})
\begin{eqnarray*}
\mathcal{H}^{\mathcal{N}}(G)= -\sum\limits_{j=1}^L\frac{V_j}{2m}
\sum\limits_{i=1}^{n_j} \frac{d_i^{(j)}}{V_j}
\log_2\frac{d_i^{(j)}}{V_j}- \sum\limits_{j=1}^L \frac{g_j}{2m}
\log_2\frac{V_j}{2m}.
\end{eqnarray*}

Set the first term of $\mathcal{H}^{\mathcal{N}}(G)$ by
$H_1=-\sum\limits_{j=1}^L\frac{V_j}{2m} \sum\limits_{i=1}^{n_j}
\frac{d_i^{(j)}}{V_j} \log_2\frac{d_i^{(j)}}{V_j}$, and for each
homochromatic set $X_j$, set $H_j=-\sum\limits_{i=1}^{n_j}
\frac{d_i^{(j)}}{V_j} \cdot \log_2\frac{d_i^{(j)}}{V_j}$. By Theorem
\ref{thm:basicproperty}, with probability $1-o(1)$, for each $j$,
$n_j\leq 4\log^{a+1}n$. Since the uniform distribution gives rise to
the maximum entropy, we have that with probability $1-o(1)$,
$$H_j \leq \log_2 n_j=O(\log\log n),$$
and by averaging, we have
$$H_1=\sum\limits_{j=1}^L \frac{V_j}{2m}
H_j=O(\log\log n).$$

Moreover,
$$-\sum_{j=1}^L \frac{g_j}{2m}\log_2\frac{V_j}{2m} \leq \sum_{j=1}^L \frac{g_j}{2m}
\log_2 2m = \frac{\log_2 2m}{2m}\cdot\sum\limits_{j=1}^L g_j.$$ Let
$m_g$ be the number of global edges in $G$. Then $\sum_{j=1}^L
g_j=2m_g$. By the construction of $G$, $m_g=d|C_n|$, where $|C_n|$
is the number of colors in $G$ (and also the number of homochromatic
sets in $G$ and the number of modules in $\mathcal{P}$). By Theorem
\ref{thm:basicproperty}, with probability $1-o(1)$,  the size
$|C_n|$ of $C_n$ is at most $2n/\log^a n$. Therefore the second term
of $\mathcal{H}^{\mathcal{N}}(G)$ is
$$-\sum_{j=1}^L \frac{g_j}{2m}\log_2\frac{V_j}{2m}
\leq \frac{\log_2 2m}{2m}\cdot\frac{4dn}{\log^a n}=O(\log^{1-a}n).$$

Putting all together, we have that, with probability $1-o(1)$,
$$\mathcal{H}^{\mathcal{N}}(G)=H_1-\sum_{j=1}^L
\frac{g_j}{2m}\log_2\frac{V_j}{2m}=O(\log\log n +\log^{1-a}n).$$ So,
if $0<a<1$, then with probability $1-o(1)$,
$$\mathcal{H}^{\mathcal{N}}(G)=O(\log^{1-a}n).$$
If $a\geq 1$, then with probability $1-o(1)$,
$$\mathcal{H}^{\mathcal{N}}(G)=O(\log\log n).$$
In both cases, $\mathcal{H}^{\mathcal{N}}(G)=o(\log n)$. Theorem
\ref{thm:resistance_security} follows.
\end{proof}

\section{Eigenvalues of the Laplacian of Resistor Graphs}\label{sec:strong resistor graph}

In this section, we study the eigenvalues of the Laplacian of the resistor graphs.
At first, we introduce some results on high order Cheeger's inequality which we will use.

Let $G=(V,E)$ be an undirected graph with $|V|=n$ and $|E|=m$. The
Laplacian of $G$ is defined to be the $n\times n$ matrix
$\mathcal{L}=I-D^{-1/2} A D^{-1/2}$, where $A$ is the adjacency
matrix of $G$ and $D$ is the diagonal matrix whose $(v,v)$-th entry
is the degree of vertex $v$. So the spectrum of $\mathcal{L}$
satisfies $0=\lambda_1\leq\lambda_2\leq\cdots\leq\lambda_n\leq 2$,
where $\lambda_k$ is the $k$-th eigenvalue of $\mathcal{L}$.

For a subset of vertices $S\subseteq V$, define the
\emph{conductance} of $S$ to be
\begin{align*}
\Phi(S)=\frac{|E(S,\overline{S})|}{\min\{\vol(S), \vol(\overline{S})\}},
\end{align*}
where $|E(S,\overline{S})|$ is the number of edges between $S$ and
its complement $\overline{S}$.

Lee, Gharan and Trevisan \cite{LGT2014} defined the \emph{$k$-way conductance} of
a graph $G$ as follows:
\begin{align*}
\phi(k)=\min\limits_{S_1,S_2,\ldots,S_k} \max\limits_{i\in [k]} \{\Phi(S_i)\},
\end{align*}
where the minimum runs over all collections of disjoint non-empty
subsets $S_1,S_2,\ldots,S_k \subseteq V$.

The high-order Cheeger's inequalities \cite{LGT2014} state that the
$k$-way conductance of $G$ is bounded by the $k$-th eigenvalue of
$\mathcal{L}$ in the following forms:
\begin{align}
\frac{\lambda_k}{2} \leq \phi(k) \leq O(k^2)\sqrt{\lambda_k},
\end{align}
\begin{align} \label{eqn:high_order_Cheeger}
\frac{\lambda_k}{2} \leq \phi(k) \leq O(\sqrt{\lambda_{2k}\cdot\log
k}).
\end{align}

In this section, we investigate the eigenvalues of the Laplacian of the graphs with optimum two-dimensional structure entropy, i.e., $\mathcal{H}^2(G)=O(\log_2\log_2n)$.

\subsection{Characterization of the graphs with small two-dimensional structure entropy}

\begin{theorem} (Combinatorial property theorem of the graphs with two-dimensional structure entropy $O(\log_2\log_2n)$) \label{thm:small_community_suf}
Let $G=(V,E)$ be a graph with number of edges $m=|E|$, volume
$\vol(G)$ and no isolated vertices. Let $w:E\rightarrow\mathbb{R}^+$ be the weight function
satisfying $\frac{\max_{e\in G} \{w(e)\}}{\min_{e\in G} \{w(e)\}}
\leq W$, for some constant $W\geq 1$. If $\mathcal{H}^2(G)\leq
c\log_2\log_2 m$ for some constant $c>0$ and any sufficiently large
$m$, then for any $\varepsilon>0$, and sufficiently large $m$, there
is a set of modules of vertices, denoted by $A$, satisfying

\begin{enumerate}
\item[(1)] $\vol(A)\geq(1-2\varepsilon)\cdot\vol(G)$;
\item[(2)] For each module $X\in A$,
$\Phi(X)\leq 1/\log_2^{1-\varepsilon}m$;
\item[(3)] For each module $X\in A$,
$|X|\leq\log^{3c/\varepsilon}m$.
\end{enumerate}
\end{theorem}

Theorem \ref{thm:small_community_suf}
implies that if $\mathcal{H}^2(G)=O(\log\log m)$, then almost all
vertices belong to a module of conductance $\varepsilon$ and size
$\log^{O(1/\varepsilon)}m$.

\begin{proof}
Since the one- and two-dimensional structural entropies depend on the
relative weights on edges instead of the absolute values of the weights, for notational
simplicity in our proof, we assume that the least weight on edge is
$1$ while the largest one is $W$. We also assume that there is no
isolated vertices in $G$ since isolated vertices do not change any
parameters in the theorem.

Let $\mathcal{P}$ be a partition of vertices in $G$ such that
$\mathcal{H}^\mathcal{P}(G)\leq c\log_2\log_2 m$. Define
$$J=\{j: V_j\in\mathcal{P}, H_j\leq\varepsilon^{-1}\cdot c\log_2\log_2 m\},$$ and
$\overline{J}=[|\mathcal{P}|]\setminus J$, where $V_j$ is the $j$-th
module of $\mathcal{P}$ and
$$H_j=-\sum\limits_{v\in
V_j}\frac{d_v}{\vol(V_j)}\log_2\frac{d_v}{\vol(V_j)}$$
\noindent is the
one-dimensional structure entropy of $V_j$. Since
$$\sum\limits_{j\in [|\mathcal{P}|]}\frac{\vol(V_j)}{\vol(G)}\cdot H_j \leq \mathcal{H}^\mathcal{P}(G) \leq c \log_2\log_2 m,$$
we have
$$\sum\limits_{j\in\overline{J}}\frac{\vol(V_j)}{\vol(G)}\cdot \varepsilon^{-1}c \log_2\log_2 m \leq
\sum\limits_{j\in\overline{J}}\frac{\vol(V_j)}{\vol(G)}\cdot H_j
\leq c\log_2\log_2 m.$$ So
$$\sum\limits_{j\in\overline{J}}\frac{\vol(V_j)}{\vol(G)} \leq
\varepsilon,$$ and
$$\sum\limits_{j\in J}\frac{\vol(V_j)}{\vol(G)} \geq
1-\varepsilon,$$ which means that the total volume of the modules in
$V_j$ for $j\in\overline{J}$, denoted by $\vol(\overline{J})$, is
negligible.

Define
$$J'=\{j: V_j\in\mathcal{P}, \Phi_j\leq\frac{1}{\log_2^{1-\varepsilon}m}\},$$ and
$\overline{J'}=[|\mathcal{P}|]\setminus J'$. Then we will show that
the total volume of the modules in $V_j$ for $j\in\overline{J'}$,
denoted by $\vol(\overline{J'})$, is also negligible. To this end,
the following lemma will be useful.

\begin{lemma} \label{lem:module_entropy}
Let $X$ be a subset of vertices in $G$ with one-dimensional structure entropy $\mathcal{H}^1(X)$ and conductance $\Phi(X)$. Let $m_X$ be
the number of edges whose two end-points are both in $X$. If $W\leq
m_X^\epsilon$ for some $\epsilon\geq 0$, then we have
$$\mathcal{H}^1(X) \geq \frac{1-\Phi(X)}{2} \cdot [(1-\epsilon)\log_2 m_X -3].$$
\end{lemma}

\begin{proof}
We call the edges whose two end-points are both in $X$ to be {\it local
edges}, and the edges in the cut $(X,\overline{X})$, each of which has
exactly one end-point in $X$, to be {\it global edges}. Let
$$g_X=\sum\limits_{e\in(X,\overline{X})} w(e)$$ be the total weight
of global edges. So $g_X=\Phi(X)\cdot\vol(X)$. Let $G_X$ be the
subgraph induced by the vertices in $X$. So $\vol(G_X)=\vol(X)-g_X$ and
the number of edges in $G_X$ is $m_X$. Since $W\leq m_X^\epsilon$,
by Theorem 21 in \cite{LP2016}, the one-dimensional
structure entropy of $G_X$ satisfies
$$\mathcal{H}^1(G_X)\geq\frac{1}{2}[(1-\epsilon)\log_2 m_X-1].$$

For each vertex $v\in X$, let $g_v$ be the total weight of global
edges associated with $v$. Let $v_i$ ($1\leq i\leq |X|$) be the
$i$-th vertex in $X$. Suppose that $v_1$ is the vertex with the largest
weighted degree in $X$ (break ties arbitrarily). Define the weighted
degree distribution
$$X'=\left\{ \frac{d_{v_0}+\sum_{i=2}^{|X|}g_{v_i}}{\vol(X)},\frac{d_{v_2}-g_{v_2}}{\vol(X)},
\ldots,\frac{d_{v_{|X|}}-g_{v_{|X|}}}{\vol(X)} \right\},$$ which is
the degree distribution obtained by associating all the global edges
to the vertex $v_1$. To establish the relationship among
$\mathcal{H}^1(X)$, $H(X')$ and $H^1(G_X)$, we introduce the
following property for Shannon entropy.

\begin{lemma} \label{lem:entropy_relation}
Let $l\geq 2$ be an integer and $\textbf{p}=\{p_1,p_2,\ldots,p_l\}$
be a probability distribution satisfying $\sum_{i=1}^l p_i=1$.
Suppose that $p_1\geq p_2$. Then for any $0\leq\alpha\leq p_2$, let
$\textbf{p}'=\{p_1+\alpha,p_2-\alpha,p_3,\ldots,p_l\}$, and we have
$H(\textbf{p})\geq H(\textbf{p}')$.
\end{lemma}

\begin{proof}
Let function
\begin{eqnarray*}
f(\alpha) &=& H(\textbf{p})-H(\textbf{p}')\\
&=&
(p_1+\alpha)\log_2(p_1+\alpha)+(p_2-\alpha)\log_2(p_2-\alpha)-p_1\log_2
p_1-p_2\log_2 p_2.
\end{eqnarray*}
So its first derivative
$$
f'(\alpha)=\log_2\frac{p_1+\alpha}{p_2-\alpha}\geq 0,
$$
for $0\leq\alpha\leq p_2$. Since $f(0)=0$, $f(\alpha)\geq 0$ for
$0\leq\alpha\leq p_2$. Lemma \ref{lem:entropy_relation} follows.
\end{proof}

Because of the symmetry of function $H(\cdot)$, Lemma
\ref{lem:entropy_relation} holds not only for $p_1$ and $p_2$, but
for any $p_i$ and $p_j$. By the construction of the distribution
$X'$, which can be viewed as associating one by one the global
edges on $v_i$ for $2\leq i\leq |X|$ to $v_1$, in which at each
step, by Lemma \ref{lem:entropy_relation}, the entropy of the
intermediate degree distribution is decreasing. Thus,
$\mathcal{H}^1(X)\geq H(X')$.

On the other hand, letting
$d_{v_1}'=d_{v_1}+\sum_{i=2}^{|X|}g_{v_i}$ and
$d_{v_i}'=d_{v_i}-g_{v_i}$ for $2\leq i\leq |X|$, by the additivity
of $H(\cdot)$, we know that

\begin{eqnarray*}
H(X') &=& H\left( \frac{d_{v_1}'}{\vol(X)},\frac{d_{v_2}'}{\vol(X)},
\ldots,\frac{d_{v_{|X|}}'}{\vol(X)} \right)\\
&=& H\left(
\frac{d_{v_1}'}{\vol(X)},\frac{\vol(X)-d_{v_1}'}{\vol(X)} \right)\\
&& + \frac{\vol(X)-d_{v_1}'}{\vol(X)} \cdot
\sum\limits_{i=2}^{|X|}\left(\frac{d_{v_i}'}{\vol(X)-d_{v_1}'}\log_2\frac{d_{v_i}'}{\vol(X)-d_{v_1}'}\right),
\end{eqnarray*}
and
\begin{eqnarray*}
H^1(G_X) &=& H\left(
\frac{d_{v_1}'-g_X}{\vol(X)-g_X},\frac{d_{v_2}'}{\vol(X)-g_X},
\ldots,\frac{d_{v_{|X|}}'}{\vol(X)-g_X} \right)\\
&=& H\left(
\frac{d_{v_1}'-g_X}{\vol(X)-g_X},\frac{\vol(X)-d_{v_1}'}{\vol(X)-g_X}
\right)\\
&& + \frac{\vol(X)-d_{v_1}'}{\vol(X)-g_X} \cdot
\sum\limits_{i=2}^{|X|}\left(\frac{d_{v_i}'}{\vol(X)-d_{v_1}'}\log_2\frac{d_{v_i}'}{\vol(X)-d_{v_1}'}\right).
\end{eqnarray*}
Comparing this two equations, we have that
\begin{align*}
&\vol(X)\cdot\left[H(X')-H\left(
\frac{d_{v_1}'}{\vol(X)},\frac{\vol(X)-d_{v_1}'}{\vol(X)}
\right)\right]\\
=& (\vol(X)-g_X)\cdot\left[\mathcal{H}^1(G_X)-H\left(
\frac{d_{v_1}'-g_X}{\vol(X)-g_X},\frac{\vol(X)-d_{v_1}'}{\vol(X)-g_X}
\right)\right],
\end{align*}
Recall that
$$\mathcal{H}^1(G_X)\geq\frac{1}{2}[(1-\epsilon)\log_2 m_X-1],$$
and $\Phi(X)=g_X/\vol(X)$, we have that
$$H(X')\geq\frac{1-\Phi(X)}{2} \cdot [(1-\epsilon)\log_2 m_X -3].$$
Recall that $\mathcal{H}^1(X)\geq H(X')$. Lemma
\ref{lem:module_entropy} follows.
\end{proof}

Then we show that $\vol(\overline{J'})$ is negligible. Assume that
there is a constant $\varepsilon_0>0$ such that
$\vol(\overline{J'})\geq\varepsilon_0\cdot\vol(G)$. Then the total
volume of the modules in $V_j$ for $j\in J'$, denoted by $\vol(J')$,
is at most $(1-\varepsilon_0)\cdot\vol(G)$ Since all the modules
$V_j$ for $j\in J'$ have a conductance at most
$1/\log_2^{1-\varepsilon}m$, then the conductance of the union of
all the modules $V_j$ for $j\in J'$ is also at most
$1/\log^{1-\varepsilon}m$ because some global edges for the modules
are possible to be local edges in the union. So the total weight of
edges in the cut $(\cup_{j\in J'}V_j,\cup_{j\in\overline{J'}}V_j)$,
denoted by $g_{\overline{J'}}$, is at most
$\vol(J')/\log_2^{1-\varepsilon}m$, which means that the conductance
of $\cup_{j\in\overline{J'}}V_j$, denoted by $\Phi(\overline{J'})$,
is at most
$\vol(J')/(\vol(\overline{J'})\cdot\log_2^{1-\varepsilon}m)\leq(1-\varepsilon_0)/(\varepsilon_0\cdot\log_2^{1-\varepsilon}m)$.
Let $m_{\overline{J'}}$ be the number of edges whose two end-points
are both in $\cup_{j\in\overline{J'}}V_j$. Recall that for each edge
$e$, $1\leq w(e)\leq W$. Then
$$m_{\overline{J'}}\geq\frac{\vol(\overline{J'})-g_{\overline{J'}}}{W}=\frac{(1-\Phi(\overline{J'}))\cdot\vol(\overline{J'})}{W}
\geq\frac{\varepsilon_0\cdot
m}{W}\cdot\left(1-\frac{1-\varepsilon_0}{\varepsilon_0\cdot\log_2^{1-\varepsilon}m}\right).$$
Since $W\leq m_{\overline{J'}}^\epsilon$ for sufficiently large $m$, by Lemma
\ref{lem:module_entropy}, the one-dimensional structure entropy of
$\cup_{j\in\overline{J'}}V_j$ satisfies
$$\mathcal{H}^1(\overline{J'}) \geq \frac{1-\Phi(\overline{J'})}{2} \cdot [(1-\epsilon)\log_2 m_{\overline{J'}} -3]=\Omega(\log m).$$
By the additivity of entropy function, we have that
$$\mathcal{H}^1(\overline{J'})=\sum\limits_{j\in\overline{J'}} \frac{\vol(V_j)}{\vol(\overline{J'})}\cdot H_j
+\sum\limits_{j\in\overline{J'}}
\left(\frac{\vol(V_j)}{\vol(\overline{J'})}\cdot\log_2\frac{\vol(V_j)}{\vol(\overline{J'})}\right)=\Omega(\log
m).$$ Since $\vol(\overline{J'})\geq\varepsilon_0\cdot\vol(G)$, we
know that $$\sum\limits_{j\in\overline{J'}}
\frac{\vol(V_j)}{\vol(G)}\cdot H_j +\sum\limits_{j\in\overline{J'}}
\frac{\vol(V_j)}{\vol(G)}\log_2\frac{\vol(V_j)}{\vol(G)}=\Omega(\log
m).$$ Note that $\Phi_j>1/\log_2^{1-\varepsilon}m$. Thus,
\begin{eqnarray*}
\mathcal{H}^\mathcal{P}(G) &=& \sum\limits_{j\in J'}
\frac{\vol(V_j)}{\vol(G)}\cdot H_j +\sum\limits_{j\in J'}
\Phi_j\cdot\frac{\vol(V_j)}{\vol(G)}\log_2\frac{\vol(V_j)}{\vol(G)}\\
&& +\sum\limits_{j\in \overline{J'}} \frac{\vol(V_j)}{\vol(G)}\cdot
H_j +\sum\limits_{j\in \overline{J'}}
\Phi_j\cdot\frac{\vol(V_j)}{\vol(G)}\log_2\frac{\vol(V_j)}{\vol(G)}\\
&=& \Omega(\log_2^\varepsilon m),
\end{eqnarray*}
which contradicts the fact that
$\mathcal{H}^\mathcal{P}(G)\leq c\log_2\log_2 m$ for some constant
$c>0$ and sufficiently large $m$. Therefore, for any
$\varepsilon>0$, $\vol(\overline{J'})\leq \varepsilon\cdot\vol(G)$
is negligible.

Now, define $A=J\cap J'$ to indicate the set of modules both in $J$ and
$J'$. So for any $j\in A$, $H_j\leq\varepsilon^{-1}\cdot
c\log_2\log_2 m$ and $\Phi_j\leq 1/\log_2^{1-\varepsilon}m$. The
total volume of modules in $A$ is at least
$\vol(G)-\vol(\overline{J})-\vol(\overline{J'})\geq(1-2\varepsilon)\cdot\vol(G)$.
So we only have to show that the size of each module $V_j$ in $A$
has a size at most $\log^{3c/\varepsilon}m$, and then Theorem
\ref{thm:small_community_suf} follows.

By Lemma \ref{lem:module_entropy}, for each $j\in A$,
$$H_j \geq \frac{1-\Phi_j}{2} \cdot [(1-\epsilon)\log_2 m_j -3],$$
if $W\leq m_j^\epsilon$, where $m_j$ is the number of edges whose
two end-points are both in $V_j$. This condition holds for any
$m_j=\omega(1)$ and in this case, $\epsilon$ can be arbitrarily
close to $0$. Otherwise, $m_j=O(1)$ and we have already shown that
the size of $V_j$ is at most $\log^{3c/\varepsilon}m$ for
sufficiently large $m$ since there is no isolated vertex in $G$.

Since $H_j\leq\varepsilon^{-1}\cdot c\log_2\log_2 m$, we have that
$$\varepsilon^{-1}\cdot c\log_2\log_2 m \geq \frac{1-\Phi_j}{2} \cdot [(1-\epsilon)\log_2 m_j -3].$$
Since $\Phi_j\leq 1/\log_2^{1-\varepsilon}m$, we have that
$$(1-\epsilon)\log_2 m_j \leq \frac{2c\log_2\log_2 m}{\varepsilon\cdot(1-1/\log_2^{1-\varepsilon}m)}+3.$$
Thus, for sufficiently large $m$, both the number of edges and the
number of vertices in $V_j$ is at most $\log^{3c/\varepsilon}m$.

This completes the proof of Theorem \ref{thm:small_community_suf}.
\end{proof}

By appropriately choosing the parameters in the proof of Theorem \ref{thm:small_community_suf}, we have the following:

\begin{theorem} \label{thm:small_community_suf_o(log)}
	Let $G=(V,E)$ be a graph with number of edges $m=|E|$ and volume
	$\vol(G)$. Let $w:E\rightarrow\mathbb{R}^+$ be the weight function
	satisfying $\frac{\max_{e\in G} \{w(e)\}}{\min_{e\in G} \{w(e)\}}
	\leq W$, for some constant $W\geq 1$. If for any $c>0$, $\mathcal{H}^2(G)\leq
	c\log_2 m$ for any sufficiently large
	$m$, then for any $\varepsilon,\phi>0$, and sufficiently large $m$, there
	is a set of modules of vertices, denoted by $A$, satisfying
	
	\begin{enumerate}
		\item[(1)] $\vol(A)\geq(1-2\varepsilon)\cdot\vol(G)$;
		\item[(2)] For each module $X\in A$,
		$\Phi(X)\leq\phi$;
		\item[(3)] For each module $X\in A$,
		$|X|\leq m^\varepsilon$.
	\end{enumerate}
\end{theorem}

Theorem \ref{thm:small_community_suf} shows that $\mathcal{H}^2(G)=O(\log\log n)$ guarantee a nice combinatorial property of the graph. On the other hand, combinatorial properties ensure that $\mathcal{H}^2(G)=O(\log\log n)$.

\begin{theorem} (Combinatorial properties guarantee $\mathcal{H}^2(G)=O(\log\log n)$) \label{thm:small_community_nec}
Let $G=(V,E)$ be a connected and balanced graph of size $n=|V|$. Then both (1) and (2) below hold:

\begin{enumerate}
\item[(1)] If there is a set of modules $A$ satisfying
\begin{enumerate}
\item[(i)] $\vol(A)=(1-o(1))\cdot\vol(G)$, where $\vol(A)$ is the sum
of the weighted degrees of all the nodes in the modules in $A$;
\item[(ii)] For each module $X\in A$, its size $|X|=n^{o(1)}$;
\item[(iii)] For each module $X\in A$, its conductance
$\Phi(X)=o(1)$,
\end{enumerate}
then the two-dimensional structural information of $G$ is
$\mathcal{H}^2(G)=o(\log n)$.

\item[(2)] If there is a set of modules $A$ satisfying
\begin{enumerate}
\item[(i)] $\vol(A)=\left(1-O\left(\frac{\log\log n}{\log n}\right)\right)\cdot\vol(G)$;
\item[(ii)] For each module $X\in A$, $|X|=\log^{O(1)}n$;
\item[(iii)] For each module $X\in A$, $\Phi(X)=O\left(\frac{\log\log n}{\log n}\right)$,
\end{enumerate}
then $\mathcal{H}^2(G)=O(\log\log n)$.
\end{enumerate}

\end{theorem}

\begin{proof}
Consider the partition of nodes of $G$ that consists of the modules
in $A$ and the module, denoted by $S$, composed by the nodes not in
$A$. Note that

\begin{eqnarray*}
\mathcal{H}^2(G) &\leq& \sum\limits_{X\in
A}\frac{\vol(X)}{\vol(G)}\cdot\mathcal{H}^1(X)+ \sum\limits_{X\in
A}\Phi(X)\cdot\frac{\vol(X)}{\vol(G)}\log_2\frac{\vol(X)}{\vol(G)}\\
&&
+\frac{\vol(S)}{\vol(G)}\cdot\mathcal{H}^1(S)+\Phi(S)\cdot\frac{\vol(S)}{\vol(G)}\log_2\frac{\vol(S)}{\vol(G)}.
\end{eqnarray*}

For (1), for each $X\in A$, we have
\begin{enumerate}
\item [(i)] $\mathcal{H}^1(X)\leq\log_2|X|=o(\log
n)$,

\item [(ii)] $\Phi(X)=o(1)$,

\item [(iii)] $\sum_{X\in
A}(\vol(X)/\vol(G))\log_2(\vol(X)/\vol(G))\leq\log_2 n$,

\item [(iv)]
$\vol(S)=o(1)\cdot\vol(G)$, $\mathcal{H}^1(S)\leq\log_2 n$, and

\item [(v)] $\Phi(S)\cdot(\vol(S)/\vol(G))\log_2(\vol(S)/\vol(G))\leq 1$.
\end{enumerate}
So in
all, $\mathcal{H}^2(G)=o(\log n)$.

For (2), for each $X\in A$, we have:

\begin{enumerate}
\item [(i)] $\mathcal{H}^1(X)\leq\log_2|X|=O(\log\log n)$,

\item [(ii)] $\Phi(X)=O(\log\log
n/\log n)$,

\item [(iii)] $\sum_{X\in
A}(\vol(X)/\vol(G))\log_2(\vol(X)/\vol(G))\leq\log_2 n$,

\item [(iv)]
$\vol(S)=O(\log\log n/\log n)\cdot\vol(G)$,

\item [(v)]
$\mathcal{H}^1(S)\leq\log_2 n$, and

\item [(vi)]
$\Phi(S)\cdot(\vol(S)/\vol(G))\log_2(\vol(S)/\vol(G))\leq 1$.

\end{enumerate}
So we have
$\mathcal{H}^2(G)=O(\log\log n)$.
\end{proof}

Theorems \ref{thm:small_community_suf} and \ref{thm:small_community_nec} together give a combinatorial characterization for the graphs $G$ with $\mathcal{H}^2(G)=O(\log\log n)$. However, in this characterization, the combinatorial property is complicated. This is the disadvantage of the combinatorial characterization theorem.
We then look for simpler characterizations of the graphs.

\subsection{Algebraic properties of the graphs with $\mathcal{H}^2(G)=O(\log\log n)$}

In this part, we show that for any connected graph $G$, if the
two-dimensional structure entropy $\mathcal{H}^2(G)=O(\log\log
n)$, then there are many eigenvalues of the Laplacian of $G$ that
are close to $0$. The proof of the result is actually an application of Theorem
\ref{thm:small_community_suf}.

\begin{theorem} \label{thm:small_community_algebraic} (Algebraic characterization theorem of the graphs with two-dimensional structure entropy $O(\log_2\log_2n)$)
For every weighted graph $G=(V,E,w)$ with number of edges $m=|E|$
and weight function $w:E\rightarrow\mathbb{R}^+$ satisfying
$\frac{\max_{e\in G} \{w(e)\}}{\min_{e\in G} \{w(e)\}} \leq W$ for
some constant $W\geq 1$, if $\mathcal{H}^2(G)\leq c\log_2\log_2 m$
for some constant $c>0$ and any sufficiently large $m$, then for any
$\varepsilon>0$ and sufficiently large $m$, there is an integer
$$k\geq
\frac{(1-2\varepsilon)\cdot m}{W\cdot\log_2^{6c/\varepsilon}m}$$ such that
$\lambda_k\leq 2/\log_2^{1-\varepsilon}m$.
\end{theorem}

 Theorem \ref{thm:small_community_algebraic} implies that if
$\mathcal{H}^2(G)=O(\log\log m)$, then there is an integer
$k=\Omega(n/\poly\log m)$ such that the $k$-th largest eigenvalue $\lambda_k$ of the
Laplacian of $G$ satisfies $\lambda_k=o(1)$.

\begin{proof}
By Theorem \ref{thm:small_community_suf}, we know that there is a
subset of vertices $A\subseteq V$ satisfying the three properties
stated in Theorem \ref{thm:small_community_suf}. Assume again that
the least weight on edge is $1$ while the largest one is $W$. For
each module $X\in A$, denote by $g_X$ the total weight of global
edges of $X$. Then we have
$$\Phi(X)=\frac{g_X}{\vol(X)}\leq\frac{1}{\log_2^{1-\varepsilon}m}.$$
Moreover, since $|X|\leq\log_2^{3c/\varepsilon}m$, we have
$$\vol(X)-g_X\leq W\cdot\log_2^{6c/\varepsilon}m.$$
Combining the two inequalities above, we have
$$\vol(X)\leq\frac{W\cdot\log_2^{6c/\varepsilon}m}{1-\log_2^{-(1-\varepsilon)}m}.$$
Since $\vol(A)\geq (1-2\varepsilon)\cdot\vol(G)$ and $\vol(G)\geq
2m$, the total number of modules in $A$ is at least
$$\dfrac{(1-2\varepsilon)\cdot 2m}{\frac{W\cdot\log_2^{6c/\varepsilon}m}{1-\log_2^{-(1-\varepsilon)}m}}
\geq\frac{(1-2\varepsilon)\cdot m}{W\cdot\log_2^{6c/\varepsilon}m}\triangleq k_0$$ for sufficiently large $m$. This means
that we can find at least $k_0$ disjoint modules in $G$, each of
which has conductance at most $1/\log_2^{1-\varepsilon}m$. By the
high-order Cheeger's inequalities, there is an integer $k\geq k_0$
such that the $k$-way conductance of $G$ is at most
$1/\log_2^{1-\varepsilon}m$, and so $\lambda_k\leq
2/\log_2^{1-\varepsilon}m$. The theorem follows.

\end{proof}

For any connected and balanced graph $G$, if $\mathcal{H}^2(G)=O(\log\log n)$, then the security index $\theta (G)$ of $G$ is $1-o(1)$. By Theorem \ref{thm:small_community_algebraic}, in this case, there is a large $k$ such that the $k$-th largest eigenvalue $\lambda_k$ of the Laplacian of $G$ is less than a small constant $\epsilon>0$. In the next section, we will show that this result can be further strengthened.

It is interesting to notice that each eigenvalue is in $[0,2]$, and that the summation of all the eigenvalues $\lambda_i$ is $n$, provided that there is no isolated vertex in the graph.
Theorem \ref{thm:small_community_algebraic} implies that the distribution of the eigenvalues of the Laplacian of a graph is closely related to the security of the graph, leading to an interesting open question to investigate the relationship between the distribution of the eigenvalues of the Laplacian of a graph and the security of the graph.

\section{Eigenvalues of the Laplacian of the Security Graphs}\label{sec:resistor graph}

%\subsection{Eigenvalues of the Laplacian of the resistor graphs}

For a given resistor graph, we first establish the following combinatorial characterization theorem.

\begin{theorem} \label{thm:small_community_suf_theta} (Combinatorial property theorem of resistor graphs)
	Let $G=(V,E)$ be a graph with number of edges $m=|E|$ and volume
	$\vol(G)$. Let $w:E\rightarrow\mathbb{R}^+$ be the weight function
	satisfying $\frac{\max_{e\in G} \{w(e)\}}{\min_{e\in G} \{w(e)\}}
	\leq W$, for some constant $W\geq 1$. If the security index $\theta(G)\geq 1-\theta$ for some constant $\theta$, then for any $\varepsilon>0$, $\phi>\theta$, there is a constant $\alpha<1$ (related to $\theta$ and $\phi$), such that for any sufficiently large $m$, there is a set of modules of vertices, denoted by $A$, satisfying
	
	\begin{enumerate}
		\item[(1)] $\vol(A)\geq(1-\alpha-\varepsilon)\cdot\vol(G)$;
		\item[(2)] For each module $X\in A$,
		$\Phi(X)\leq\phi$;
		\item[(3)] For each module $X\in A$,
		$|X|\leq 2^{\mathcal{H}^1(G)\cdot\frac{3\theta}{\varepsilon(1-\phi)(1-\epsilon)}}$.
	\end{enumerate}
\end{theorem}

\begin{proof}
	We also suppose that the edge weights range from $1$ to $W$. Since $\theta(G)\geq 1-\theta$, there is a partition $\mathcal{P}$ on $V$ such that $\mathcal{H}^\mathcal{P}(G)\leq\theta\cdot\mathcal{H}^1(G)$. Define
	$$J=\{j:V_j\in\mathcal{P},H_j\leq\varepsilon^{-1}\cdot\theta\mathcal{H}^1(G)\},$$
	where $V_j$ is the $j$-th module of $\mathcal{P}$ and
	$$H_j=-\sum\limits_{v\in V_j}\frac{d_v}{\vol(V_j)}\log_2\frac{d_v}{\vol(V_j)}$$ is the one-dimensional structure entropy of $V_j$. Thus the fraction of total volume of $\overline{J}$ satisfies
	$$\sum\limits_{j\in \overline{J}}\frac{\vol(V_j)}{\vol(G)} \leq
	\varepsilon.$$
	
	Define
	$$J'=\{j: V_j\in\mathcal{P}, \Phi_j\leq\phi\},$$
and
	$\overline{J'}=[|\mathcal{P}|]\setminus J'$. Then we will show that
	the fraction of the total volume of the modules in $V_j$ for $j\in\overline{J'}$,
	denoted by $\vol(\overline{J'})$, is at most some constant $\alpha$. Let $\vol(\overline{J'})=\varepsilon_0\cdot\vol(G)$. Then $\mathcal{H}^\mathcal{P}(G)\leq\theta\cdot\mathcal{H}^1(G)$ means that
	$$(1-\theta)\cdot\sum\limits_{j\in[|\mathcal{P}|]}\frac{\vol(V_j)}{\vol(G)}\cdot H_j \leq (\theta-\Phi_j)\cdot\hat{H}(J')+(\theta-\Phi_j)\cdot\hat{H}(\overline{J'}),$$
	where $$\hat{H}(J')=-\sum\limits_{j\in J'}\frac{\vol(V_j)}{\vol(G)}\log_2\frac{\vol(V_j)}{\vol(G)}$$ and $$\hat{H}(\overline{J'})=-\sum\limits_{j\in \overline{J'}}\frac{\vol(V_j)}{\vol(G)}\log_2\frac{\vol(V_j)}{\vol(G)}.$$
	By the definition of $J'$, we have
	$$(1-\theta)\cdot\sum\limits_{j\in[|\mathcal{P}|]}\frac{\vol(V_j)}{\vol(G)}\cdot H_j \leq \theta\cdot\hat{H}(J')+(\theta-\phi)\cdot\hat{H}(\overline{J'}).$$
	Replacing $\vol(J')$ and $\vol(\overline{J'})$ with $\varepsilon_0\cdot\vol(G)$ and $(1-\varepsilon_0)\cdot\vol(G)$, respectively, we get
	$$(1-\theta)\cdot\sum\limits_{j\in[|\mathcal{P}|]}\frac{\vol(V_j)}{\vol(G)}\cdot H_j \leq \theta(1-\varepsilon_0)\cdot H(J')+(\theta-\phi)\varepsilon_0\cdot H(\overline{J'})-\theta(1-\varepsilon_0)\log_2(1-\varepsilon_0)-(\theta-\phi)\varepsilon_0\log_2\varepsilon_0,$$
	where $$H(J')=-\sum\limits_{j\in J'}\frac{\vol(V_j)}{\vol(J')}\log_2\frac{\vol(V_j)}{\vol(J')}$$ and $$H(\overline{J'})=-\sum\limits_{j\in \overline{J'}}\frac{\vol(V_j)}{\vol(\overline{J'})}\log_2\frac{\vol(V_j)}{\vol(\overline{J'})}.$$
	This implies that
	$$(1-\theta)\cdot\sum\limits_{j\in[|\mathcal{P}|]}\frac{\vol(V_j)}{\vol(G)}\cdot H_j \leq \theta\cdot H(J')-[\theta\cdot H(J')+(\phi-\theta)\cdot H(\overline{J'})]\cdot\varepsilon_0+2\theta.$$
	
	Since the left hand side is non-negative, $\varepsilon_0$ is not negligible unless $H(\overline{J'})$ is negligible compared with $H(J')$. In the latter case, if $H(\overline{J'}) \leq \varepsilon'\cdot H(J')$ for a small enough $\varepsilon'$, by a straightforward calculation, we have
	$$\hat{H}(\overline{J'})=\frac{\varepsilon'\varepsilon_0}{1-\varepsilon_0}\cdot\hat{H}(J')+(1-\varepsilon_0)\log_2(1-\varepsilon_0)-\varepsilon_0\log_2\varepsilon_0.$$
	So if $1-\varepsilon_0=o(1)$, then both $\hat{H}(\overline{J'})$ and $\hat{H}(J')$ are $o(\log n)$, which contradicts to the fact that $\hat{H}(\overline{J'})+\hat{H}(J') \geq (1-\theta)\cdot\mathcal{H}^1(G)$. Therefore, there is a constant $\alpha<1$ such that $\varepsilon_0\leq\alpha$.
	
	So define $A=J\cap\overline{J'}$, and thus for any $j\in A$, $H_j\leq\varepsilon^{-1}\cdot
	\theta\mathcal{H}^1(G)$ and $\Phi_j\leq\phi$. The
	total volume of modules in $A$ is at least
	$\vol(G)-\vol(\overline{J})-\vol(\overline{J'})\geq(1-\varepsilon-\alpha)\cdot\vol(G)$.
	So the only task is to show that the size of each such module is at most $2^{\mathcal{H}^1(G)\cdot\frac{3\theta}{\varepsilon(1-\phi)(1-\epsilon)}}$.
	
	By Lemma \ref{lem:module_entropy}, for each $j\in A$,
	$$\varepsilon^{-1}\cdot\theta\mathcal{H}^1(G)\geq H_j\geq\frac{1-\Phi_j}{2} \cdot [(1-\epsilon)\log_2 m_j -3].$$
	Since $\Phi_j\leq\phi$, we have that $$(1-\epsilon)\log_2 m_j\leq\frac{2\theta\cdot\mathcal{H}^1(G)}{\varepsilon(1-\phi)}+3.$$
	Therefore, for sufficiently large $m$, both the number of edges $m_j$ and the
	number of vertices in $V_j$ is upper bounded by $2^{\mathcal{H}^1(G)\cdot\frac{3\theta}{\varepsilon(1-\phi)(1-\epsilon)}}$. This completes the proof of Theorem \ref{thm:small_community_suf_theta}.
\end{proof}

By Theorem \ref{thm:small_community_suf_theta}, we have the following algebraic property theorem of the general resistor graphs.

\begin{theorem} \label{thm:small_community_algebraic_theta} (Algebraic property theorem of resistor graphs)
	For every weighted graph $G=(V,E,w)$ with number of edges $m=|E|$
	and weight function $w:E\rightarrow\mathbb{R}^+$ satisfying
	$\frac{\max_{e\in G} \{w(e)\}}{\min_{e\in G} \{w(e)\}} \leq W$ for
	some constant $W\geq 1$, if the security index $\theta(G)\geq 1-\theta$ for some constant $\theta$, then for any $\varepsilon>0$, $\phi>\theta$, there is a constant $\alpha<1$ such that for any sufficiently large $m$, there is an integer
	$$k\geq
	\frac{2(1-\alpha-\varepsilon)(1-\phi)\cdot m}{W\cdot 2^{\mathcal{H}^1(G)\cdot\frac{6\theta}{\varepsilon(1-\phi)(1-\epsilon)}}}$$ such that
	$\lambda_k\leq 2\phi$.
\end{theorem}
\begin{proof}
By the proof of Theorem \ref{thm:small_community_algebraic}.
\end{proof}

%\subsection{Algebraic characterisation of the security graphs}
%
%Let $C$ be the constant implicit in the upper bound of Eq. (\ref{eqn:high_order_Cheeger}), that is,
%\begin{align} \label{eqn:high_order_Cheeger_C}
%\frac{\lambda_k}{2} \leq \phi(k) \leq C\cdot\sqrt{\lambda_{2k}\cdot\log
%	k}
%\end{align}
%for some constant $C>0$. Then the following theorem implies that a large set of small eigenvalues of the Laplacian indicates high security index.
%
%\begin{theorem} \label{thm:characterisation-of-security-graph}
%	(Characterisation of the resistor graphs by eigenvalues of the Laplacian of the graphs) Let $G=(V,E)$ be a graph with number of edges $m=|E|$ and volume $\vol(G)$. Let $w:E\rightarrow\mathbb{R}^+$ be the weight function
%	satisfying $\frac{\max_{e\in G} \{w(e)\}}{\min_{e\in G} \{w(e)\}}
%	\leq W$, for some constant $W\geq 1$. If there is $k\geq C\cdot n^{1-\varepsilon}$ such that the $2k$-th largest eigenvalue of the Laplacian of $G$ satisfies $\lambda_{2k}\leq 1/(n^{2\varepsilon}\cdot\log_2 k)$, then the security index $\theta (G)$ of $G$ is at least $1-2\varepsilon$.
%\end{theorem}
%
%\begin{proof}
%\end{proof}

\section{Conclusions and Discussion}\label{sec:con}

We proposed the notion of resistance of a graph to measure the force of the graph to resist cascading failures of strategic virus attacks. The resistance of a graph $G$ is the maximum number of bits required to determine the codeword of the module of the graph that is accessible from random walk from which random walk cannot escape. We found the resistance law of networks that the resistance of a graph is the difference of the one- and two-dimensional structure entropy of the graph. Here, for a graph $G$ and a natural number $K$, the $K$-dimensional structure entropy of $G$ is the least number of bits required to determine the $K$-dimensional codeword of the vertex that is accessible from the random walk with stationary distribution in $G$. We defined the security index of a graph $G$ to be the normalised resistance of $G$. We propose the notion of $(n,\theta)$-resistor graph. For a large constant $\theta$ (that is, less than and close to $1$), an $(n,\theta)$-resistor graph is a connected graph with $n$ vertices, and with security index greater than or equal to $\theta$.
We showed that for a tree with bounded weights or grid graphs $G$, the resistance of $G$ is $\Omega (\log n)$ and the security index of $G$ is $1-o(1)$. The results demonstrate that the natural structures such as trees and grid graphs have the important property of high resistance and high security against virus attacks. We showed that for the networks $G$ of the security model with affinity exponent $a>0$ and edge parameter $d\geq 2$, the resistance of $G$ is maximised as $\Omega (\log n)$, and the security index of $G$ is maximised as $1-o(1)$, for sufficiently large $n$. Therefore, the security model does generate the networks of high resistances and high security indices. We also establish both a combinatorial and an algebraic characterization theorems of the resistor graphs. In particular, we show that for a large constant $\theta$, for an $(n,\theta)$-resistor graph, and for any small constant $\epsilon>0$, there is a large $k$ such that the $k$-th largest eigenvalue of the Laplacian of the graph is less than $\epsilon$. Our results provide the fundamental theory for network security, with potential applications in the security engineering of networks.

%\newpage

\section*{Appendix A: Probabilistic Tools}

We will use the following form of Chernoff bound.

\begin{lemma}\label{lem:chernoff} (Chernoff bound, \cite{C1981}) Let $X_1,\ldots,X_n$ be independent random variables
with $\Pr[X_i=1]=p_i$ and $\Pr[X_i=0]=1-p_i$. Denote the sum by
$X=\sum\limits_{i=1}^n X_i$ with expectation
$E(X)=\sum\limits_{i=1}^n p_i$. Then we have
$$\Pr[X\leq E(X)-\lambda]\leq \exp\left(-\frac{\lambda^2}{2E(X)}\right),$$
$$\Pr[X\geq E(X)+\lambda]\leq \exp\left(-\frac{\lambda^2}{2(E(X)+\lambda/3)}\right).$$
\end{lemma}

We will use the following form of Azuma's inequality for
martingales.

\begin{lemma}\label{lem:Azuma} (Azuma's inequality)
Let $\textbf{c}=(c_1,\ldots,c_n)$ be a vector of positive entries.
Let a sequence of random variables $X_0,X_1,\ldots,X_n$ be a
martingale. If it is $\textbf{c}$-Lipschitz, that is,
$|X_i-X_{i-1}|\leq c_i$ for $i=1,\ldots,n$, then for any
$\lambda>0$,
$$\Pr[X_n\leq X_0-\lambda]\leq \exp\left(-\frac{\lambda^2}{2\sum_{i=1}^n c_i^2}\right),$$
$$\Pr[X_n\geq X_0+\lambda]\leq \exp\left(-\frac{\lambda^2}{2\sum_{i=1}^n c_i^2}\right).$$
\end{lemma}

We will use the following form of supermartingale inequality.

\begin{lemma}\label{lem:supermartingale} (Supermartingale inequality, \cite{CL2006} Theorem 2.40)
For a filter
$\{0,\Omega\}=\mathcal{F}_0\subset\mathcal{F}_1\subset\cdots\subset\mathcal{F}_n=\mathcal{F}$,
suppose that a non-negative random variable $X_i$ is
$\mathcal{F}_i$-measurable for $0\leq i\leq n$. Let $B$ be the bad
set associated with the following admissible conditions: (that is,
the set of events that the conditions fail to hold.)
\begin{eqnarray*}
E(X_i | \mathcal{F}_{i-1}) &\leq& X_{i-1},\\
\Var(X_i | \mathcal{F}_{i-1}) &\leq& \sigma_i^2 +\phi_i X_{i-1},\\
X_i-E(X_i | \mathcal{F}_{i-1}) &\leq& a_i + M,
\end{eqnarray*}
where $\sigma_i$, $\phi_i$, $a_i$ and $M$ are non-negative
constants. Then we have
\begin{eqnarray*}\Pr(X_n \geq X_0+\lambda)
&\leq& \exp\left( -\frac{\lambda^2}{2(\sum_{i=1}^n
(\sigma_i^2+a_i^2)+(X_0+\lambda)(\sum_{i=1}^n \phi_i)+M\lambda/3)}
\right)+\Pr(B).
\end{eqnarray*}
\end{lemma}

The following fact will also be very useful in our proofs.

\begin{fact}\label{fac:log}
For any real $x$,
$$\frac{1}{x+1} \leq \log\left(1+\frac{1}{x}\right) \leq \frac{1}{x}.$$
\end{fact}

\begin{proof}
Note that $1+y \leq e^{y}$ holds for all real $y$. The fact is
obtained by replacing $y$ with $-\frac{1}{x+1}$ and $\frac{1}{x}$,
respectively.
\end{proof}

The following expansion of power series is folklore.

\begin{fact}\label{fac:expansion}
For any $u>0$ and $|x|\leq 1$,
\begin{eqnarray*}
(1\pm x)^u &=& 1\pm
ux+\frac{u(u-1)}{2!}x^2\pm\frac{u(u-1)(u-2)}{3!}x^3\\
&& +\cdots+(-1)^m\frac{u(u-1)\cdots(u-m+1)}{m!}x^m+\cdots.
\end{eqnarray*}
\end{fact}

\section*{Appendix B: Proof of Theorem \ref{thm:basicproperty}}

\begin{proof}(Proof of Theorem \ref{thm:basicproperty})
For (1). By the construction of $G$, the expectation of $|C_t|$ is
$$E[|C_t|] = n_0+ \sum_{i=3}^{t}\frac{1}{\log^a i}.$$
By indefinite integral
$$\int (\frac{1}{\log^a x}-\frac{a}{\log^{a+1}x}) dx=\frac{x}{\log^a x}+C,$$
we know that if $t\geq T_1$ is large enough (when $n$ is large
enough), then
\begin{eqnarray*}
\sum_{i=3}^t \frac{1}{\log^a i} & \leq &
1+\int_2^t \frac{1}{\log^a x}dx\\
& \leq & \int_2^t \frac{6}{5}(\frac{1}{\log^a
x}-\frac{a}{\log^{a+1}x}) dx\\
& \leq & \frac{4t}{3\log^a t},
\end{eqnarray*}
where $\frac{6}{5}$ and $\frac{4}{3}$ are chosen arbitrarily among
the numbers larger than $1$. Similarly,
\begin{eqnarray*}
\sum_{i=3}^t \frac{1}{\log^a i} & \geq &
\int_2^t \frac{1}{\log^a x}dx\\
& \geq & \int_2^t (\frac{1}{\log^a
x}-\frac{a}{\log^{a+1}x}) dx\\
& \geq & \frac{3t}{4\log^a t}.
\end{eqnarray*}

By the Chernoff bound (Lemma \ref{lem:chernoff}), since $t\geq T_1$
and $n_0$ is a constant, with probability
$1-exp(-\Omega(\frac{t}{\log^a t}))=1-o(n^{-1})$, we have
$\frac{t}{2\log^a t}\leq |C_t|\leq \frac{2t}{\log^a t}$. By the
union bound, such an inequality holds for all $t\geq T_1$ with
probability $1-o(1)$.

we define the following event:

\begin{definition}\label{def:eventE}
Let $\mathscr{E}$ be the event that, for all $i\geq T_1$,
$\frac{i}{2\log^a i}\leq |C_i|\leq \frac{2i}{\log^a i}$.
\end{definition}

By the discussion above, $\mathscr{E}$ happens with probability
$1-o(1)$. We will assume and use this event frequently throughout
our proofs.

\bigskip

For (2). By the construction of $G$, the expectation of $|S|$ at
time step $t$ is
$$E(|S|)=1+\sum\limits_{i=t_S+1}^t \left(1-\frac{1}{\log^a i}\right) \cdot \frac{1}{|C_i|}.$$

By (1), we know that $\mathscr{E}$ holds with probability $1-o(1)$.
Thus, if $a>0$, then at time step $t$,
\begin{eqnarray*}
E(|S|) &=& \Theta\left(\sum\limits_{i=t_S}^t \left(1-\frac{1}{\log^a
t}\right) \cdot
\frac{\log^a t}{t}\right)\\
&=& \Theta\left( \int_{t_S}^t \frac{\log^a x}{x}dx \right)\\
&=& \Theta(\log^{a+1}t-\log^{a+1}t_S).
\end{eqnarray*}

\bigskip

For (3). It suffices to show that with probability $1-o(n^{-1})$,
the homochromatic set of the first color $\kappa$ has size
$4\log^{a+1}n$. Then the result follows from the union bound.

Let $S_\kappa$ be the set of vertices sharing color $\kappa$. Conditioned
on the event $\mathscr{E}$, for large enough $n$,

\begin{eqnarray*}
E(|S_\kappa|) &=& 1+\sum\limits_{i=3}^n \left(1-\frac{1}{\log^a
i}\right) \cdot
\frac{1}{|C_i|}\\
&\leq& T_1+\sum\limits_{i=T_1+1}^n \left(1-\frac{1}{\log^a i}\right)
\cdot \frac{2\log^a i}{i}\\
&\leq& 3\log^{a+1}n.
\end{eqnarray*}

By the Chernoff bound,
$$\Pr[|S_\kappa|>4\log^{a+1} n] = o(n^{-1}).$$
Therefore, with probability $1-o(n^{-1})$, the size of $S_\kappa$ is
at most $4\log^{a+1}n$.

\bigskip

For (4). We need to bound the number of global edges with one
endpoint in $S$.

For $t\geq t_S$, define $S[t]$ to be the snapshot of $S$ at time
step $t$, and $\partial(S)[t]$ to be the set of edges from $S[t]$ to
$\overline{S[t]}$, the complement of $S[t]$. So $\partial(S)[t]$ is
in fact the set of global edges of $S$ at time step $t$ and
$g_S=|\partial(S)[n]|$. Denote by $D(S)[t]$ the total degree of
vertices in (the volume of) $S[t]$. In our proof, we first give a
recurrence for the expected value of $D(S)[t]$ at any time step
$t>t_S$, and then show that $\partial(S)[n]$ is not expectedly too
many.

By the construction of $G$, when a new vertex is created, the volume
it contributes to the network is $2d$. By (1), we know that the
volume of $G_t$ is $2d(1+o(1))t$, where $o(t)$ is contributed by
$G_{n_0}$. The recurrence of $D(S)[t]$ satisfies
\begin{eqnarray} \label{eqn:D_recurrence}
E[D(S)[t]\ |\ D(S)[t-1]] &\leq& D(S)[t-1]+\frac{1}{\log^a t}
\left[ \frac{D(S)[t-1]}{2d(t-1)}+(d-1)\cdot\frac{1}{|C_{t-1}|} \right] \nonumber \\
 && +\left(1-\frac{1}{\log^a t}\right)\cdot\frac{2d}{|C_{t-1}|}.
\end{eqnarray}

We suppose the event $\mathscr{E}$ that for all $t\geq
T_1=\log^{a+1}n$, $\frac{t}{2\log^a t}\leq |C_t|\leq
\frac{2t}{\log^a t}$, which almost surely holds by (1). It also
holds for $t\geq T_2$ for sufficiently large $n$. On this condition,
recalling that $d\geq 2$, we have
\begin{eqnarray} \label{eqn:degree}
E[D(S)[t]\ |\ D(S)[t-1]]&\leq& D(S)[t-1]
\left[1+\frac{1}{2(t-1)\log^a t}\right] +\frac{4d\log^a t}{t}.
\end{eqnarray}
Taking expectation on both sides, we have
\begin{eqnarray}
E(D(S)[t]) \leq E(D(S)[t-1]) \left[1+\frac{1}{2(t-1)\log^a t}\right]
+ \frac{4d\log^a t}{t}. \label{eqn:exp_degree}
\end{eqnarray}
Then we analyze this recurrence for the cases of $a\geq 1$ and
$a<1$, respectively.

\smallskip

When $a\geq 1$, since for sufficiently large $n$ and thus for
sufficiently large $t$ with $t\geq t_S\geq T_2$, we have
\begin{eqnarray}
&&9d\log^{a+1}(t+1)-\left[1+\frac{1}{2(t-1)\log^a t}\right] \cdot
9d\log^{a+1} t \nonumber \\
&\geq& 9d\log^a t\log\frac{t+1}{t}-\frac{9d\log
t}{2(t-1)} \nonumber \\
&\geq& \frac{9d\log^a t}{t+1}-\frac{9d\log^a t}{2(t-1)} \nonumber \\
&\geq& \frac{4d\log^a t}{t}, \label{eqn:adjust_a>=1}
\end{eqnarray}
where the second inequality follows from Fact \ref{fac:log}.
Applying it to Inequality (\ref{eqn:exp_degree}), we have
\begin{eqnarray*}
E(D(S)[t])-9d\log^{a+1}(t+1) \leq \left[1+\frac{1}{2(t-1)\log^a
t}\right] \cdot (E(D(S)[t-1])-9d\log^{a+1}t).
\end{eqnarray*}
Recursively, we have
\begin{eqnarray*}
E(D(S)[t]) \leq \theta_t \cdot
[E(D(S)[t_S])-9d\log^{a+1}(t_S+1)]+9d\log^{a+1}(t+1)
\end{eqnarray*}
holds for all $t_S<t\leq n$, where
\begin{eqnarray*}
\theta_t=\prod\limits_{i=t_S+1}^t \left[1+\frac{1}{2(i-1)\log^a
i}\right].
\end{eqnarray*}
Note that $E(D(S)[t_S])=d$. So
\begin{eqnarray}
E(D(S)[t]) \leq 9d\log^{a+1}(t+1)-\theta_t \cdot
[9d\log^{a+1}(t_S+1)-d]. \label{eqn:exp_degree_a>=1}
\end{eqnarray}

\smallskip

When $0<a<1$, since for sufficiently large $n$ and thus for
sufficiently large $t$,
\begin{eqnarray}
&&\left[1+\frac{1}{2(t-1)\log^a t}\right] \cdot
9d\log^{2a}t-9d\log^{2a}(t+1) \nonumber \\
&=& \frac{9d\log^a t}{2(t-1)}-9d \cdot [\log^{2a}(t+1)-\log^{2a}t]\nonumber  \\
&\geq& \frac{9d\log^a t}{2(t-1)}-\frac{d\log^a t}{2t} \nonumber \\
&\geq& \frac{4d\log^a t}{t}, \label{eqn:adjust_a<1}
\end{eqnarray}
where the first inequality follows from the fact that
$\log(t+1)-\log t=\log\left( 1+\frac{1}{t} \right)\leq \frac{1}{t}$
and so when $a<1$,
\begin{eqnarray*}
\lim\limits_{t\rightarrow\infty} \frac{\log^{2a}
(t+1)-\log^{2a}t}{\frac{\log^a t}{t}} &=&
\lim\limits_{t\rightarrow\infty} t \cdot
\left[\frac{\log^a(t+1)}{\log^a t}-1\right] \cdot (\log^a(t+1)+\log^a t) \\
&\leq& \lim\limits_{t\rightarrow\infty} t \cdot
\left[\frac{\log(t+1)}{\log t}-1\right] \cdot (\log^a(t+1)+\log^a t) \\
&\leq& \lim\limits_{t\rightarrow\infty} t \cdot \frac{\log(t+1)-\log
t}{\log t} \cdot 2\log^a(t+1)\\
&\leq& \lim\limits_{t\rightarrow\infty} \frac{2\log^a(t+1)}{\log t}
=0.
\end{eqnarray*}

Applying Inequality (\ref{eqn:adjust_a<1}) to
(\ref{eqn:exp_degree}), we have
\begin{eqnarray*}
E(D(S)[t])+9d\log^{2a}(t+1) \leq \left[1+\frac{1}{2(t-1)\log^a
t}\right] \cdot (E(D(S)[t-1])+9d\log^{2a}t).
\end{eqnarray*}
Recursively, we have
\begin{eqnarray*}
E(D(S)[t]) \leq \theta_t \cdot
[E(D(S)[t_S])+9d\log^{2a}(t_S+1)]-9d\log^{2a}(t+1)
\end{eqnarray*}
holds for all $t_S<t\leq n$, and so
\begin{eqnarray}
E(D(S)[t]) \leq \theta_t \cdot [9d\log^{2a}(t_S+1)+d]-
9d\log^{2a}(t+1). \label{eqn:exp_degree_a<1}
\end{eqnarray}

\smallskip

Note that by the construction of $G$,
\begin{eqnarray} \label{eqn:exp_g_S}
E(g_S)\leq\sum\limits_{t=t_S}^n \frac{1}{\log^a t} \left[ \frac{
E(D(S)[t])}{2d(t-1)}+E\left(\frac{d-1}{|C_{t-1}|}\right) \right].
\end{eqnarray}
Let $$U_1=\sum\limits_{t=t_S}^n \frac{E(D(S)[t])}{2d(t-1)\log^a t}$$
and $$U_2=\sum\limits_{t=t_S}^n
E\left(\frac{d-1}{|C_{t-1}|\cdot\log^a t}\right).$$ So $E(g_S)\leq
U_1+U_2$. Recall that in the proof of (1), we have shown that for
each time step $t\geq T_1$($\geq T_2$), with probability at least
$1-\exp(-\Omega(\frac{t}{\log^a t}))$, we have $\frac{t}{2\log^a
t}\leq |C_t|\leq \frac{2t}{\log^a t}$. So for some constant $c>0$,
$$E\left(\frac{d-1}{|C_{t-1}|\cdot\log^a t}\right)\leq
\frac{2(d-1)}{t}+t\cdot\exp\left(-\frac{ct}{\log^a t}\right),$$ and
so
$$U_2\leq \sum\limits_{t=t_S}^n E\left(\frac{d-1}{|C_{t-1}|\cdot\log^a t}\right)=O(d\cdot(\log n-\log t_S))=O(d\cdot\log\log n).$$
Next, we will bound $U_1$ by using Inequalities
(\ref{eqn:exp_degree_a>=1}) and (\ref{eqn:exp_degree_a<1}) for
different values of $a$.

\smallskip

When $a\geq 1$, we have
$$U_1\leq \sum\limits_{t=t_S}^n \frac{9d\log^{a+1}(t+1)-\theta_t\cdot[9d\log^{a+1}(t_S+1)-d]}{2d(t-1)\log^a t}.$$
Since $\theta_t>1$, for sufficient large $n$, we have
\begin{eqnarray*}
U_1 &\leq& \sum\limits_{t=t_S}^n
\frac{9\log^{a+1}(t+1)-[9\log^{a+1}(t_S+1)-1]}{2(t-1)\log^a t}\\
&\leq& \frac{9}{2} \left[\sum\limits_{t=t_S}^n \frac{\log
t}{t-1}-\log^{a+1}(t_S+1) \sum\limits_{t=t_S}^n
\frac{1}{(t-1)\log^a t} \right]\\
&\leq& \frac{9}{2} \left( \int_{t_S}^n \frac{\log x}{x}dx -
\log^{a+1}t_S \int_{t_S}^n \frac{1}{x\log^a x}dx \right).
\end{eqnarray*}

If $a>1$, then
\begin{eqnarray*}
U_1 &\leq& \frac{9}{2} \cdot\left[ \frac{1}{2}(\log^2 n-\log^2
t_S)-\frac{\log^{a+1}t_S}{1-a}(\log^{1-a} n-\log^{1-a} t_S))
\right]\\
&=& \frac{9}{2} \log^2 n \cdot\left[
\frac{1}{2}-\left(\frac{1}{2}+\frac{1}{a-1}\right) \left(\frac{\log
t_S}{\log n}\right)^2 + \frac{1}{a-1}\left(\frac{\log t_S}{\log
n}\right)^{a+1} \right]\\
&=& \frac{9}{2} \log^2 n \cdot\left[ \frac{1}{2}-\frac{a+1}{2(a-1)}
\left(1-\frac{b\log\log n}{\log n}\right)^2 +
\frac{1}{a-1}\left(1-\frac{b\log\log n}{\log n}\right)^{a+1}
\right].
\end{eqnarray*}
By Fact \ref{fac:expansion},
$$\left(1-\frac{b\log\log n}{\log n}\right)^{a+1}\leq 1-\frac{(a+1)b\log\log n}{\log n}+\frac{(a+1)ab^2(\log\log n)^2}{2\log^2n}.$$
Thus,
\begin{eqnarray*}
U_1 &\leq& \frac{9}{2} \log^2 n \cdot \left[
\frac{1}{2}-\frac{a+1}{2(a-1)} \left(1-\frac{2b\log\log n}{\log
n}+\frac{b^2(\log\log n)^2}{\log^2
n}\right) \right. \\
&& \left. + \frac{1}{a-1}\left(1-\frac{(a+1)b\log\log n}{\log
n}+\frac{(a+1)ab^2(\log\log n)^2}{2\log^2n}\right) \right]\\
&=& \frac{9}{4} (a+1)b^2(\log\log n)^2.
\end{eqnarray*}
Note that $E(g_S)=U_1+U_2$ and $U_2=O(\log\log n)$. For sufficiently
large $n$, $E(g_S)\leq \frac{5}{2}(a+1)b^2(\log\log n)^2$. (4)(i)
follows.

If $a=1$, then
\begin{eqnarray*}
U_1 &\leq& \frac{9}{2} \cdot\left( \int_{t_S}^n \frac{\log x}{x}dx -
\log^2 t_S \int_{t_S}^n \frac{1}{x\log x}dx \right)\\
&=& \frac{9}{2} \left[ \frac{1}{2}(\log^2 n-\log^2 t_S)-\log^2 t_S
\cdot
(\log\log n-\log\log t_S) \right]\\
&=& \frac{9}{2} \left[ \frac{1}{2}(\log^2 n-\log^2 t_S)-\log^2 t_S
\cdot
\log\left(1+\frac{b\log\log n}{\log n-b\log\log n}\right) \right]\\
&\leq& \frac{9}{2} \left[ \frac{1}{2}(\log^2 n-\log^2 t_S)-\log^2
t_S \cdot
\frac{b\log\log n}{\log n} \right]\\
&=& \frac{9}{2} \left[ \frac{1}{2}\log^2 n-\frac{1}{2}(\log
n-b\log\log n)^2-(\log n-b\log\log n)^2 \cdot
\frac{b\log\log n}{\log n} \right]\\
&=& \frac{9}{2} \left[ \frac{3}{2}b^2(\log\log n)^2-\frac{(b\log\log
n)^3}{\log n} \right]\\
&\leq& \frac{27}{4} b^2(\log\log n)^2.
\end{eqnarray*}
Since $E(g_S)=U_1+U_2$ and $U_2=O(\log\log n)$, when $n$ is large
enough, $E(g_S)\leq 8b^2(\log\log n)^2$. (4)(ii) follows.

When $a<1$, applying Inequality (\ref{eqn:exp_degree_a<1}) to
(\ref{eqn:exp_g_S}), we have
\begin{eqnarray*}
U_1 &\leq& \sum\limits_{t=t_S}^n \frac{\theta_t \cdot
(9d\log^{2a}(t_S+1)+d)- 9d\log^{2a}(t+1)}{2d(t-1)\log^a t}\\
&\leq& \frac{9}{2} \cdot \left[ \sum\limits_{t=t_S}^n
\frac{\theta_n\log^{2a}t_S}{(t-1)\log^a t}-\sum\limits_{t=t_S}^n
\frac{log^{2a}(t+1)}{(t-1)\log^a t} \right]\\
&=& \frac{9}{2} \cdot \left( \theta_n\log^{2a}t_S \cdot \int_{t_S}^n
\frac{1}{x\log^a x}dx-\int_{t_S}^n
\frac{log^a x}{x}dx \right)+O\left(\frac{1}{n}\right)\\
&=& \frac{9}{2} \cdot \left( \theta_n\log^{2a}t_S \cdot
\frac{\log^{1-a}n-\log^{1-a}t_S}{1-a}-\frac{\log^{1+a}n-
\log^{1+a}t_S}{1+a} \right)+O\left(\frac{1}{n}\right)\\
&=& \frac{9\theta_n}{2(1-a)}\log^{1-a}n\log^{2a}t_S
-\frac{9}{2}\left(\frac{\theta_n}{1-a}-\frac{1}{1+a}\right)
\log^{1+a}t_S\\
&&-\frac{9}{2(1+a)}\log^{1+a}n+O\left(\frac{1}{n}\right)\\
&=& \frac{9\theta_n}{2(1-a)}\log^{1+a}n\left(1-\frac{b\log\log
n}{\log
n}\right)^{2a}-\frac{9}{2}\left(\frac{\theta_n}{1-a}-\frac{1}{1+a}\right)
\log^{1+a}n\\
&& \cdot\left(1-\frac{b\log\log
n}{\log n}\right)^{1+a}-\frac{9}{2(1+a)}\log^{1+a}n+O\left(\frac{1}{n}\right)\\
&=& \frac{9\theta_n}{2(1-a)}\log^{1+a}n \cdot \left[
1-\frac{2ab\log\log n}{\log
n}+\frac{2a(2a-1)}{2}\left(\frac{b\log\log n}{\log
n}\right)^2\right.\\
&& \left.+O\left(\frac{\log\log n}{\log n}\right)^3 \right]
-\frac{9}{2}\left(\frac{\theta_n}{1-a}-\frac{1}{1+a}\right)\log^{1+a}n
\cdot \left[ 1-\frac{(1+a)b\log\log n}{\log
n}\right.\\
&& \left.+\frac{(a+1)a}{2}\left(\frac{b\log\log n}{\log n}\right)^2
+O\left(\frac{\log\log n}{\log n}\right)^3
\right]-\frac{9}{2(1+a)}\log^{a+1}n+O\left(\frac{1}{n}\right)\\
&=& \left[ \frac{9\theta_n}{2(1-a)}
-\frac{9}{2}\left(\frac{\theta_n}{1-a}-\frac{1}{1+a}\right)
-\frac{9}{2(1+a)} \right]\cdot\log^{1+a}n +\left[
-\frac{9}{2(1-a)}\cdot 2ab\right.\\
&& \left.+\frac{9}{2}
\left(\frac{\theta_n}{1-a}-\frac{1}{1+a}\right)\cdot(1+a)b
\right]\cdot\log^a n \log\log n +O\left[\frac{(\log\log
n)^2}{\log^{1-a}n}\right]\\
&=& \frac{9}{2} b(\theta_n-1)\log^a n\log\log
n+O\left[\frac{(\log\log n)^2}{\log^{1-a}n}\right].
\end{eqnarray*}

To deal with the factor $(\theta_n-1)$, we need the following lemma.

\begin{lemma}\label{lem:theta}
For sufficiently large $n$, $$\theta_n-1\leq \frac{b\log\log
n}{\log^a n}.$$
\end{lemma}
Note that by the above lemma, for sufficiently large $n$
\begin{eqnarray*}
U_1 &\leq& \frac{9}{2} b\cdot\frac{b\log\log n}{\log^a n}\log^a
n\log\log n+O\left[\frac{(\log\log
n)^2}{\log^{1-a}n}\right]\\
&\leq& \frac{9}{2} b^2 (\log\log n)^2.
\end{eqnarray*}
Note that $E(g_S)=U_1+U_2$ and $U_2=O(\log\log n)$. For sufficiently
large $n$, $E(g_S)\leq 5b^2 (\log\log n)^2$. (4)(iii) follows.

To complete the proof, we prove Lemma \ref{lem:theta}.

\begin{proof}
Recall that
$$\theta_n=\prod\limits_{i=t_S+1}^n \left[1+\frac{1}{2(i-1)\log^a i}\right].$$
Then
\begin{eqnarray*}
\log\theta_n &=& \sum\limits_{i=t_S+1}^n
\log\left[1+\frac{1}{2(i-1)\log^a i}\right]\\
&\leq& \sum\limits_{i=t_S+1}^n \frac{1}{2(i-1)\log^a i}\\
&\leq& \frac{1}{2} \int_{t_S}^n \frac{1}{x\log^a x}\\
&=& \frac{1}{2(1-a)}\cdot(\log^{1-a}n-\log^{1-a}t_S)\\
&=& \frac{\log^{1-a}n}{2(1-a)}\cdot\left[1-\left(1-\frac{b\log\log
n}{\log n}\right)^{1-a}\right]\\
&=& \frac{\log^{1-a}n}{2(1-a)}\cdot\left[(1-a)\cdot\frac{b\log\log
n}{\log n}
-\frac{(1-a)(-a)}{2}\cdot\left(\frac{b\log\log n}{\log n}\right)^2\right.\\
&& \left.+O\left(\frac{\log\log n}{\log n}\right)^3\right]\\
&=& \frac{b\log\log n}{2\log^a n}+O\left[\frac{(\log\log
n)^2}{\log^{1+a}n}\right].
\end{eqnarray*}
Thus, for sufficiently large $n$, $\log\theta_n\leq\frac{3b\log\log
n}{4\log^a n}$, which implies that
$$\theta_n\leq(\log n)^\frac{3b}{4\log^a n}.$$
A key observation is that, for any constant $c$, by l'H\^{o}pital's
rule,
\begin{align*}
\lim\limits_{n\rightarrow\infty}\frac{(\log n)^\frac{c}{\log^a
n}-1}{\frac{\log\log n}{\log^a n}} &=
\lim\limits_{y\rightarrow\infty} \frac{y^\frac{c}{y^a}-1}{\frac{\log
y}{y^a}} = \lim\limits_{y\rightarrow\infty}
\frac{\left(y^\frac{c}{y^a}-1\right)'}{\left(\frac{\log
y}{y^a}\right)'}\\
&= \lim\limits_{y\rightarrow\infty} \frac{c(1-a\log
y)}{y^{1+a-\frac{c}{y^a}}} \cdot \frac{y^{1+a}}{1-a\log y}\\
&= \lim\limits_{y\rightarrow\infty} c\cdot y^\frac{c}{y^a} =
\lim\limits_{y\rightarrow\infty} c\cdot e^\frac{c\log y}{y^a} = c.
\end{align*}
Thus, for any $\epsilon>0$, if $n$ is large enough, then
$$\theta_n-1 \leq \frac{3b}{4}(1+\epsilon)\cdot\frac{\log\log n}{\log^a n}.$$
Let $\epsilon=\frac{1}{3}$, then the lemma follows.
\end{proof}

This completes the proof of Theorem \ref{thm:basicproperty}.
\end{proof}

\end{document}